\documentclass[letterpaper, aps, prd, twocolumn, superscriptaddress, showpacs, nofootinbib]{revtex4}
\pdfoutput=1

\usepackage{graphicx}
\usepackage{bm}
\usepackage{amssymb}
\usepackage{amsmath}
\usepackage{amsfonts}
\usepackage{dcolumn}
\usepackage{natbib}
\usepackage{color}
\usepackage[latin9]{inputenc}
\usepackage{url}

\sloppypar

\newcommand{\raiseentry}[1]{\smash{\raise 0.7 em \hbox{#1}}}

\newcommand{\scri}{\ensuremath{\mathcal{J}}}
\newcommand{\news}{\ensuremath{\mathcal{N}}}

\newenvironment{equationarray}
{\arraycolsep 0.14 em
\begin{eqnarray}}
{\end{eqnarray}}

\newenvironment{equationarray*}
{\arraycolsep 0.14 em
\begin{eqnarray*}}
{\end{eqnarray*}}

\newcommand{\code}[1]{{\tt #1}}
\newcommand{\change}[1]{{\color{red} #1}}

\begin{document}

\title{Gravitational Wave Extraction in Simulations of Rotating
  Stellar Core Collapse}

\author{C. Reisswig}
\email{reisswig@tapir.caltech.edu}
\affiliation{TAPIR, MC 350-17, California Institute of Technology, 1200 E California Blvd.,
Pasadena, CA 91125, USA}

\author{C. D.\ Ott}
\email{cott@tapir.caltech.edu}
\affiliation{TAPIR, MC 350-17, California Institute of Technology, 1200 E California Blvd.,
Pasadena, CA 91125, USA}
\affiliation{Center for Computation \& Technology, Louisiana State University, 216 
Johnston Hall, Baton Rouge, LA 70803, USA}

\author{U. Sperhake}
\email{sperhake@tapir.caltech.edu}
\affiliation{Institut de Ci\`encies de l'Espai (CSIC-IEEC), Facultat de Ci\`encies, Campus UAB, E-08193 Bellaterra, Spain}
\affiliation{TAPIR, MC 350-17, California Institute of Technology, 1200 E California Blvd.,
Pasadena, CA 91125, USA}
\affiliation{Department of Physics and Astronomy, The University of Mississippi, University, MS 38677-1848, USA}

\author{E. Schnetter}
\email{schnetter@cct.lsu.edu}
\affiliation{Center for Computation \& Technology, Louisiana State
  University, 216 Johnston Hall, Baton Rouge, LA 70803, USA}
\affiliation{Department of Physics \& Astronomy,
  Louisiana State University, 202 Nicholson Hall, Baton Rouge, LA 70803, USA}
\affiliation{Perimeter Institute, 31 Caroline St.\ N., Waterloo,
   Ontario N2L 2Y5, Canada}

\date{\today}


\begin{abstract}
  We perform simulations of general relativistic rotating stellar core
  collapse and compute the gravitational waves (GWs) emitted in the
  core bounce phase of three representative models via multiple
  techniques. The simplest technique, the quadrupole formula (QF),
  estimates the GW content in the spacetime from the mass quadrupole
  tensor only.  It is strictly valid only in the weak-field and
  slow-motion approximation.  For the first time, we apply GW
  extraction methods in core collapse that are fully curvature-based
  and valid for strongly radiating and highly relativistic sources.
  These techniques are not restricted to weak-field and slow-motion
  assumptions.  We employ three extraction methods computing (i) the
  Newman-Penrose (NP) scalar $\Psi_4$, (ii)
  Regge-Wheeler-Zerilli-Moncrief (RWZM) master functions, and (iii)
  Cauchy-Characteristic Extraction (CCE) allowing for the extraction
  of GWs at future null infinity, where the spacetime is
  asymptotically flat and the GW content is unambiguously defined.
  The latter technique is the only one not suffering from residual
  gauge and finite-radius effects.  All curvature-based methods suffer
  from strong non-linear drifts. We employ the fixed-frequency
  integration technique as a high-pass waveform filter.  Using the CCE
  results as a benchmark, we find that finite-radius NP extraction
  yields results that agree nearly perfectly in phase, but differ in
  amplitude by $\sim 1-7\%$ at core bounce, depending on the model.
  RWZM waveforms, while in general agreeing in phase, contain spurious
  high-frequency noise of comparable amplitudes to those of the
  relatively weak GWs emitted in core collapse.  We also find
  remarkably good agreement of the waveforms obtained from the QF with
  those obtained from CCE.  The results from QF agree very well in
  phase and systematically underpredict peak amplitudes by
  $\sim5-11\%$, which is comparable to the NP results and is certainly
  within the uncertainties associated with core collapse physics.
\end{abstract}


\maketitle


\section{Introduction}
\label{section:introduction}

Massive stars ($M \gtrsim 8-10\,M_\odot$) end their nuclear burning
lives with a core composed primarily of iron-group nuclei embedded in
an onion-skin structure of progressively lighter elements.  
Energy generation has ceased in such a star's high-density core and
relativistically-degenerate electrons provide pressure support against
gravity. Silicon shell burning, neutrino cooling, and deleptonization
eventually push the core over its effective
Chandrasekhar mass.  Radial instability sets in, leading to core
collapse, accelerated by electron capture and photodisintegration of
iron-group nuclei (see, e.g., \cite{bethe:90,baron:90}).

The collapsing iron core separates into a subsonically collapsing
homologous ($v\propto r$) inner core and supersonically infalling
outer core. When the former reaches nuclear density, the nuclear
equation of state (EOS) stiffens, dramatically increasing central
pressure support and stabilizing the inner core, which, due to its
large inertia, overshoots its new equilibrium and then rebounds
into the still collapsing outer core, launching the hydrodynamic
supernova shock. 
The acceleration experienced by the inner core in this \emph{core
  bounce} is tremendous, leading to the reversal of the collapse
velocities of order $0.1 c$ of its $\sim 0.5\, M_\odot$ of material on
a millisecond timescale. 

It was realized early on that the large accelerations encountered in stellar
collapse in combination with a source of quadrupole (or higher) order
asphericity lead to the emission of a burst of gravitational waves
(GWs; see \cite{ott:09} for a historical overview). Rotation,
centrifugally deforming the inner core to oblate shape, is an obvious
source of such quadrupole asymmetry and rotating core collapse and
bounce is the most extensively studied GW emission process in stellar
collapse~(see, e.g.,
\cite{ott:07prl,dimmelmeier:08,scheidegger:10,scheidegger:10b,takiwaki:10,ott:11a}
for recent studies and references therein). Alternatively, asymmetries
in collapse may arise from perturbations, e.g., due to large
convective plumes in the final phase of core nuclear burning, and may
lead to GW emission at bounce and/or seed GW-emitting prompt
postbounce convection~\cite{bh:96,fryer:04,ott:09}. A multitude of GW
emission processes may be active in the postbounce, pre-explosion
phase. These include convection/turbulence in the protoneutron star
and in the postshock region, nonaxisymmetric rotational instabilities
of the protoneutron star, protoneutron star pulsations, instabilities
of the standing accretion shock, and asymmetric emission of neutrinos
(\cite{ott:09,ott:06prl,marek:09,murphy:09,kotake:09,yakunin:10} and
references therein).

Of the entire ensemble of potential GW emission processes in stellar
collapse, rotating core collapse and bounce is arguably the simplest
and yields the cleanest signal, depending only on rotation, on the
nuclear EOS, and on the mass of the inner core at bounce
\cite{dimmelmeier:08}. Moreover, 3D studies have shown that collapsing
iron cores with rotation rates in the range of what is physically
plausible stay axisymmetric throughout the collapse phase and develop
nonaxisymmetric dynamics only after
bounce~\cite{shibata:05,ott:07prl,scheidegger:10}.  Hence, the GW
signal of rotating core collapse and bounce is linearly polarized and
axisymmetric (2D) simulations are sufficient for its prediction.
Unlike postbounce dynamics involving large scale and small scale fluid
instabilities of stochastic nature, the GW signal of rotating collapse
and bounce can, in principle, be predicted exactly for a given set of
initial data. Hence, it has the potential of being used in GW searches
using matched-filtering techniques (e.g., \cite{thorne:87}) or
alternative approaches also taking into account detailed signal
predictions~\cite{roever:09,summerscales:08}.

Much progress has been made in recent years in the modeling of
rotating core collapse and its GW signature. State-of-the-art
simulations are general relativistic (GR)
\cite{dimmelmeier:02,dimmelmeier:05,shibata:04,shibata:05,
  shibata:06,ott:07prl,ott:07cqg,kuroda:10,dimmelmeier:07,
  dimmelmeier:08} and some studies include magnetic fields
\cite{shibata:06,obergaulinger:06b,kuroda:10} or finite-temperature
EOS, deleptonization, and progenitors from stellar evolutionary
calculations \cite{ott:07prl,ott:07cqg,dimmelmeier:07,
  dimmelmeier:08}. These improvements in the physics included in core
collapse models provide for a more accurate and reliable dynamics
underlying the emission of GWs. The calculation of the GW signal
itself, however, is still being carried out predominantly in the
slow-motion, weak-field quadrupole approximation (e.g.,
\cite{thorne:80}) that is of questionable quality, given the extreme
densities and velocities involved in core collapse.  The quadrupole
formula (QF) ``extracts'' GWs based on matter dynamics alone, is not
invariant under general relativistic gauge transformations, treats the emission region
as a point source, and suffers from the fact that the definition of
the generalized mass quadrupole moment is not unique in GR.

In GR, the GW content of a spacetime can be extracted by means of the
perturbative Regge-Wheeler-Zerilli-Moncrief (RWZM)
formalism~\cite{Regge1957,Zerilli1970b,Zerilli:1971wd,Moncrief74}
which is gauge invariant to first order or via the Newman-Penrose (NP)
scalars approach~\cite{Newman1962,Penrose1963} which depends on the
non-unique choice of the tetrad in which the Newman-Penrose scalars
are evaluated. For reliable results, both RWZM and NP require
extraction in the wave zone \cite{thorne:80} at coordinate radii many
wavelengths from the source, but even there, coordinate ambiguities
exist. The latter are removed only when GWs are extracted at future
null infinity ($\scri^+$, see \cite{Newman1962,Penrose1963}), where space
is asymptotically flat.

Shibata \& Sekiguchi~\cite{shibatasekiguchi:03} have used simulations
of an oscillating polytropic neutron star model to compare QF and
finite-radius RWZM results. For the same basic system,
Baiotti~et~al.~\cite{baiotti:09} compared QF, finite-radius RWZM, and
finite-radius NP GW extraction with each other and with results from a
1D perturbation analysis. Both studies found that in the context of neutron 
star oscillations, the \emph{phase} of the waveforms obtained with
the quadrupole approximation agrees exceptionally well with that of the RWZM and NP
extraction methods. Shibata \&
Sekiguchi, using their particular choice of the generalized
quadrupole moment, found a systematic $\sim 20\%$ underprediction of
the GW \emph{amplitudes} by the QF\@. Baiotti~et~al.~\cite{baiotti:09}, who
studied multiple incarnations of the QF, found either underprediction
or overprediction of the amplitude, both by up to $\sim60\%$, depending on the
particular choice of QF\@.  Nagar~et~al.~\cite{nagar:05b} studied the
performance of RWZM and QF-based GW extraction from oscillating
polytropic tori and found qualitatively similar results, and
quantitative differences in amplitudes and integrated emitted energies
$E_\mathrm{GW}$ between $\sim 2\%$ and $\sim 25\%$, again depending on the
choice of quadrupole moment.

RWZM and NP GW extraction and comparisons with the QF approximation
for GWs emitted in core collapse spacetimes have proven difficult. On
the one hand, the emitted GWs are weak: Typical strain amplitudes are
$Dh \sim 10-1000\,\mathrm{cm}$, where D is the distance to the source,
and typical emitted energies are of order $10^{-10}-10^{-8}\,M_\odot
c^2$ \cite{ott:09}, many orders of magnitude lower than what is
expected, for example, from double neutron star coalescence
\cite{oechslin:07} or binary black hole mergers \cite{Reisswig:2009vc,Lousto:2009mf}. 
On the other hand, the GWs have typical
frequencies of $100 - 1000\,\mathrm{Hz}$ and corresponding wavelengths of
$300 - 3000\,\mathrm{km}$, hence require extraction at large
coordinate radii where the grid resolution of core collapse
simulations is typically too low to allow extraction of the relatively
low-amplitude GWs emitted in core collapse (see, e.g., the discussion
in \cite{ott:06phd}). Shibata \& Sekiguchi, in \cite{shibata:05}, were
able to extract GWs with the RWZM formalism from an extreme core
collapse model that developed a rotationally-induced large-scale
nonaxisymmetric deformation after bounce, emitting GWs with $Dh \sim
20000\,\mathrm{cm}$. For this model, they found that the QF accurately
predicts the GW phase, but underestimates the strain amplitude by $\sim 10\%$. 
Due to the aforementioned difficulties, these
authors were unable to compare RWZM with QF for more moderate,
axisymmetric models. Cerd\'a-Dur\'an~et~al.~\cite{cerda:05} performed
core collapse simulations using a second-order post-Newtonian (2PN)
extension of the conformal-flatness approximation to GR\@. Exploiting an
approximate relationship of the non-conformal 2PN part of the metric
to its GW part \cite{cerda:05}, they were able to extract GWs from
their 2PN metric in standard axisymmetric rotating core collapse
models. They found very close agreement (to a few percent in strain
amplitude) between QF and 2PN GW signals for almost all considered
collapse models. Siebel \textit{et al.}~\cite{Siebel03} performed nonrotating axisymmetric core collapse simulations by employing
evolutions based on a fully general relativistic null cone formalism. They added nonspherical perturbations to the star,
leading to the emission of GWs which they were able to extract with the Bondi news function at $\mathcal{J}^+$.
Comparisons to the QF suggested a significant discrepancy in amplitude and frequency from the more reliable
Bondi news result.

The results of Shibata \& Sekiguchi~\cite{shibata:05} and of
Cerd\'a-Dur\'an~et~al.~\cite{cerda:05} provide some handle on the
performance of the QF approximation in core collapse spacetimes. The
former study, while being performed in full GR, considered only a
single extreme model. In addition, the authors were forced to extract
GWs with RWZM at too small radii for completely reliable results.  The
latter study, while considering a broader ensemble of models, was
restricted to 2PN without considering full GR, leaving room for doubts
about the quality of their GW extraction technique.
Finally, the results of Siebel \textit{et al.}~\cite{Siebel03} were limited to axisymmetry without rotation
and are unreliable in the presence of strong shocks~\cite{Siebel03}.

In this study, we readdress GW extraction from rotating core collapse
spacetimes. We perform $3+1$ GR hydrodynamics simulations of rotating
core collapse, for the first time in the core collapse context
extracting GWs with RWZM, NP, and multiple QFs and comparing the
results of these methods. In addition, and also for the first time in
the present context, we utilize the Cauchy-Characteristic Extraction
(CCE) approach~\cite{Winicour05, Bishop97b, Reisswig:2009us,
  Reisswig:2009rx, Babiuc:2010ze} that propagates the GW information
to $\scri^+$ for completely gauge independent and unambiguous GW
extraction.  

In choosing our models set, we are guided by
Cerd\'a-Dur\'an~et~al.~\cite{cerda:05}, and draw precollapse
configurations from the set of \cite{dimmelmeier:02}.  These models
are GR $n=3$-polytropic iron cores in rotational equilibrium and we
evolve them with an analytic hybrid polytropic/$\Gamma$-law EOS used
in many previous studies of rotating core
collapse~\cite{zwerger:97,dimmelmeier:02,
  obergaulinger:06b,shibata:04,shibata:05,shibata:06,cerda:05}.  For
physically accurate GW signal predictions to be used in GW data
analysis, a microphysically more complete treatment is warranted.
Fortunately, recent results of studies employing such modeling technology
(e.g., \cite{ott:07prl,ott:07cqg, dimmelmeier:07,
  dimmelmeier:08,abdikamalov:10}) show that, with a proper choice of
EOS parameters, hybrid-EOS models are able to qualitatively and
to some extent quantitatively reproduce the GW signals obtained
with the much more complex and computationally intensive microphysical
studies. Hence, for the purpose of this study, we resort to the
simpler hybrid-EOS models.

Our simulations employ the open-source {\tt Zelmani} GR core collapse
simulation package~\cite{ott:09c} that is based on the {\tt Cactus
  Computational Toolkit} \cite{goodale:03, cactusweb} and the {\tt
  Einstein Toolkit} \cite{einsteintoolkitweb}. While using the full
$3+1$ GR formalism, we limit our simulations to an octant of the 3D
cube, using periodic boundary conditions on two of the inner faces of
the octant and reflective boundary conditions on the third face. This
limits 3D structure to even $\ell$ and $m$ that are multiples of $4$, 
which is not a limitation for the current study, since rotating
core collapse and the very early postbounce evolution are likely to
proceed nearly
axisymmetrically~\cite{ott:07prl,scheidegger:08,scheidegger:10}.
We note that, even though the GW signal in rotating core collapse is
dominated by the $(\ell=2,m=0)$ '+' polarization mode, there is no
reason to expect different behavior for other GW multipoles or
polarizations and our results should translate to the non-axisymmetric
case.

The results of our simulations indicate that NP extraction yields results
that agree well with those obtained from the most sophisticated CCE method. 
We observe differences in amplitude of $1-7\%$, depending on the model,
while the agreement in phase is nearly perfect.
We also find that the RWZM formalism yields unphysical high-frequency signal
components that make this method less suitable for core collapse simulations
where the signal is very weak.
Finally, we note that the quadrupole approximation
yields surprisingly close results to those obtained from CCE\@.
While the phases  nearly perfectly agree, the amplitude shows differences of $5-11\%$.

This paper is structured as follows. In Sec.~\ref{sec:methods}, we
discuss our methodology, initial data, and EOS
details. Section~\ref{sec:waveextract} discusses the various GW extraction
methods that we employ. In Sec.~\ref{sec:results}, we present our results
and discuss them in detail. Finally, in Sec.~\ref{sec:summary}, we summarize
and review our findings.


\section{Methods}
\label{sec:methods}

We adopt the ADM $ 3 + 1 $ foliation of spacetime~\cite{york:79}.  All
equations assume $c = G = 1 $ unless noted otherwise. In the
following, Latin indices run from 1 to 3 while Greek ones run from 0
to 3. We adhere to abstract index notation. $ g_{\mu\nu} $ is the
4-metric, $ \gamma_{ij} $ is the 3-metric
and $K_{ij}$ the extrinsic curvature.

\subsection{Infrastructure and Mesh Refinement: Cactus and Carpet}
\label{sec:amr}

Our code uses the {\tt Cactus Computational Toolkit} \cite{goodale:03,
  cactusweb} to manage the complexity inherent in large software
projects. {\tt Cactus} is an open source high-performance computing
environment designed for scientists and engineers; its modular
structure enables parallel computation across different architectures,
and facilitates collaborative code development between different
groups. Indeed, our code uses a set of components of the public
{\tt EinsteinToolkit} \cite{ES-Schnetter2008n, einsteintoolkitweb}, a
community project developing and supporting open software for
relativistic astrophysics, such as e.g.\ the curvature and
hydrodynamics evolution methods described below. Many improvements
made in the course of the research for this paper were contributed
back to the community.

In particular, {\tt Cactus} allows us to clearly separate between physics
components and computational components in our code. Distributed
memory parallelism in {\tt Cactus} is provided by a \emph{driver} component
which implements the data structures discretising the manifold on
which the computational state vector lives. In our case, this is the
{\tt Carpet} driver \cite{ES-Schnetter2003b, ES-Schnetter2006a,
  ES-carpetweb} providing block-structured adaptive mesh refinement
(AMR) and multi-block discretization. {\tt Carpet} parallelizes using a
hybrid approach combining MPI and OpenMP, where inter-node
communication is handled via MPI and intra-node communication via
OpenMP or via MPI, depending on the particular system and on details
of the simulation setup.

{\tt Carpet} implements Berger-Oliger style AMR \cite{Berger1984}, where the
fine grids are aligned with coarse grids, refined by factors of two.
{\tt Carpet} also implements subcycling in time, where finer grids take two
time steps for every coarse grid step. The latter greatly improves
efficiency, but also introduces significant complexity into the time
evolution method. The refined regions can be chosen and modified
arbitrarily, which we use here to add additional, finer levels during
evolution as successively higher resolutions are required to capture
the collapse. This is described in more detail in \cite{ott:06phd}.

We use fifth-order accurate spatial interpolation for spacetime
variables and third order essentially non-oscillatory interpolation for
hydrodynamics variables. Time interpolation, which is necessary to
provide boundary conditions to fine levels at times where there is no
coarse level, is second-order accurate. We apply no time refinement
between levels 3 and 4, which corresponds to reducing the Courant-Friedrichs-Lewy factor
on levels 3 and coarser by a factor of 2. This increases the accuracy
on level 3 where we extract gravitational waves.
In total, we use 9 refinement levels (including the base grid), an outer
boundary radius of $3840 M_\odot$ ($\sim 5700\,\mathrm{km}$) and a finest
zone size of $0.25 M_\odot$ $(\sim 370\,\mathrm{m})$ in our baseline resolution.

\subsection{Curvature Evolution: McLachlan}
\label{sec:curv}

\subsubsection{Evolution System}

We evolve the full Einstein equations in a $3+1$ split (a Cauchy
initial boundary value problem) \cite{York1979}, using the BSSN
formulation \cite{Alcubierre2000}, a $1+\log$ slicing
\cite{Alcubierre2003b}, and $\Gamma$-driver shift condition
\cite{Alcubierre2003b}. This leads to the following set of evolved
variables:
\begin{eqnarray}
  \phi & := & \log \left[ \frac{1}{12} \det \gamma_{ij} \right]
  \\
  \tilde\gamma_{ij} & := & e^{-4\phi}\; \gamma_{ij}
  \\
  K & := & g^{ij} K_{ij}
  \\
  \tilde A_{ij} & := & e^{-4\phi} \left[ K_{ij} - \frac{1}{3} g_{ij} K
    \right]
  \\
  \tilde\Gamma^i & := & \tilde\gamma^{jk} \tilde\Gamma^i_{jk} .
\end{eqnarray}
Our exact evolution equations are as described by Eqs.~(3) to (10) of
\cite{ES-Brown2007b}, which we list here for completeness:
\begin{widetext}
\begin{eqnarray}
  \partial_0 \alpha & = & -\alpha^2 f(\alpha, \phi, x^\mu) (K -
  K_0(x^\mu))
  \\
  \partial_0 K & = & -e^{-4\phi} \left[ \tilde{D}^i \tilde{D}_i \alpha
    + 2 \partial_i \phi \cdot \tilde{D}^i \alpha \right] + \alpha
  \left( \tilde{A}^{ij} \tilde{A}_{ij} + \frac{1}{3} K^2 \right) - \alpha S
  \\
  \partial_0 \beta^i & = & \alpha^2 G(\alpha,\phi,x^\mu) B^i
  \\
  \partial_0 B^i & = & e^{-4\phi} H(\alpha,\phi,x^\mu)
  \partial_0\tilde{\Gamma}^i - \eta^i(B^i,\alpha,x^\mu)
  \\
  \partial_0 \phi & = & -\frac{\alpha}{6}\, K +
  \frac{1}{6}\partial_k\beta^k
  \\
  \partial_0 \tilde{\gamma}_{ij} & = & -2\alpha\tilde{A}_{ij} 
  + 2\tilde{\gamma}_{k(i}\partial_{j)}\beta^k 
  - \frac{2}{3}\tilde{\gamma}_{ij}\partial_k\beta^k
  \\
  \partial_0 \tilde{A}_{ij} & = & e^{-4\phi}\left[ 
    \alpha\tilde{R}_{ij} + \alpha R^\phi_{ij} - \tilde{D}_i\tilde{D}_j\alpha 
    + 4\partial_{(i}\phi\cdot\tilde{D}_{j)}\alpha\right]^{TF}
  \nonumber\\
  & & {} + \alpha K\tilde{A}_{ij} - 2\alpha\tilde{A}_{ik}\tilde{A}^k_{\; j}
  + 2\tilde{A}_{k(i}\partial_{j)}\beta^k 
  - \frac{2}{3}\tilde{A}_{ij}\partial_k\beta^k
  - \alpha e^{-4\phi} \hat{S}_{ij}
  \\
  \partial_0\tilde{\Gamma}^i & = & 
  \tilde{\gamma}^{kl}\partial_k\partial_l\beta^i
  + \frac{1}{3} \tilde{\gamma}^{ij}\partial_j\partial_k\beta^k 
  + \partial_k\tilde{\gamma}^{kj} \cdot \partial_j\beta^i
  - \frac{2}{3}\partial_k\tilde{\gamma}^{ki} \cdot \partial_j\beta^j
  \nonumber\\
  & & {} - 2\tilde{A}^{ij}\partial_j\alpha 
  + 2\alpha\left[ (m-1)\partial_k\tilde{A}^{ki} - \frac{2m}{3}\tilde{D}^i K
    + m(\tilde{\Gamma}^i_{\; kl}\tilde{A}^{kl} + 6\tilde{A}^{ij}\partial_j\phi)
    \right] - S^i
\end{eqnarray}
\end{widetext}
with the momentum constraint damping constant set to $m=1$. The stress
energy tensor $T_{\mu\nu}$ is incorporated via the projections
\begin{eqnarray}
  \rho & := & \frac{1}{\alpha^2} \left( T_{00} - 2 \beta^i T_{0i} +
  \beta^i \beta^j T^{ij} \right)
  \\
  S & := & g^{ij} T_{ij}
  \\
  S_i & := & - \frac{1}{\alpha} \left( T_{0i} - \beta^j T_{ij} \right) .
\end{eqnarray}
We have introduced the notation $\partial_0 = \partial_t -
\beta^j\partial_j$. All quantities with a tilde $\tilde{~}$ refer to
the conformal 3-metric $\tilde{\gamma}_{ij}$, which is used to
raise and lower indices. In particular, $\tilde{D}_i$ and
$\tilde{\Gamma}^k_{ij}$ refer to the covariant derivative and the
Christoffel symbols with respect to $\tilde{\gamma}_{ij}$. The
expression $[ \cdots ]^{TF}$ denotes the trace-free part of the
expression inside the parentheses, and we define the Ricci tensor
contributions
\begin{widetext}
\begin{eqnarray}
\tilde{R}_{ij} 
 &=& -\frac{1}{2} \tilde{\gamma}^{kl}\partial_k\partial_l\tilde{\gamma}_{ij} 
  + \tilde{\gamma}_{k(i}\partial_{j)}\tilde{\Gamma}^k
  - \tilde{\Gamma}_{(ij)k}\partial_l\tilde{\gamma}^{lk} 
  + \tilde{\gamma}^{ls}\left( 2\tilde{\Gamma}^k_{\; l(i}\tilde{\Gamma}_{j)ks} 
  + \tilde{\Gamma}^k_{\; is}\tilde{\Gamma}_{klj} \right)
\\
R^\phi_{ij} &=& -2\tilde{D}_i\tilde{D}_j\phi 
  - 2\tilde{\gamma}_{ij} \tilde{D}^k\tilde{D}_k\phi
  + 4\tilde{D}_i\phi\, \tilde{D}_j\phi 
  - 4\tilde{\gamma}_{ij}\tilde{D}^k\phi\, \tilde{D}_k\phi .
\end{eqnarray}
\end{widetext}
This is a so-called $\phi$-variant of BSSN. The evolved gauge
variables are lapse $\alpha$, shift $\beta^i$, and a quantity $B^i$
related to the time derivative of the shift. The gauge parameters $f$,
$G$, $H$, and $\eta$ are determined by our choice of a $1+\log$
slicing:
\begin{eqnarray}
  f(\alpha,\phi,x^\mu) & := & 2/\alpha
  \\
  K_0(x^\mu) & := & 0
\end{eqnarray}
and $\Gamma$-driver shift condition:
\begin{eqnarray}
  G(\alpha,\phi,x^\mu) & := & (3/4)\, \alpha^{-2}
  \\
  H(\alpha,\phi,x^\mu) & := & \exp\{4\phi\}
  \\\label{eq:eta}
  \eta(B^i,\alpha,x^\mu) & := & (1/2)\, B^i q(r) .
\end{eqnarray}
The expression $q(r)$ attenuates the $\Gamma$-driver depending on the
radius as described below.

The $\Gamma$-driver shift condition is symmetry-seeking,
driving the shift $\beta^i$ to a state that renders the conformal 
connection functions $\tilde\Gamma^i$
stationary. Of course, such a stationary state cannot be achieved
while the metric is evolving, but in a stationary spacetime the time
evolution of the shift
$\beta^i$ and thus that of the spatial coordinates $x^i$ will be exponentially
damped. This damping time scale is set by the gauge parameter $\eta$
(see Eq.~\ref{eq:eta}) which has dimension $1/T$ (inverse time).
As described, e.g., in \cite{Muller:2009jx, ES-Schnetter2010a}, this
time scale may need to be adapted in different regions of the domain
to avoid spurious high-frequency behavior in regions that otherwise
evolve only very slowly, e.g., far away from the source.

Here we use the simple damping mechanism described in Eq.~(12) of
\cite{ES-Schnetter2010a}, which is defined as
\begin{eqnarray}
  \label{eq:varying-simple}
  q(r) & := & \left\{
    \begin{array}{llll}
      1 & \mathrm{for} & r \le R & \textrm{(near the origin)}
      \\
      R/r & \mathrm{for} & r \ge R & \textrm{(far away)}
    \end{array}
    \right.
\end{eqnarray}
with a constant $R$ defining the transition radius between the
interior, where $q\approx1$, and the exterior, where $q$ falls off as
$1/r$. Eq.~\ref{eq:eta} describes how $q$ appears in the gauge
parameters. In this paper we use $R=250\,M_\odot$ ($R =
369.2\,\mathrm{km}$).

We implement the above BSSN equations and gauge conditions in the
{\tt McLachlan} code \cite{ES-Brown2007b, ES-mclachlanweb} which is
freely available as part of the {\tt EinsteinToolkit}. {\tt McLachlan} is
auto-generated from the definition of the variables and equations in the
{\tt Mathematica} format by the {\tt Kranc} code generator \cite{kranc04,
  Husa:2004ip, krancweb}. {\tt Kranc} is a suite of {\tt Mathematica} packages
comprising a computer algebra toolbox for numerical relativists. {\tt Kranc}
can be used as a ``rapid prototyping'' system for physicists or
mathematicians handling complex systems of partial differential
equations, and through integration into the {\tt Cactus} framework one can
also produce efficient production codes.

We use fourth-order accurate finite differencing for the spacetime
variables and add a fifth-order Kreiss-Oliger dissipation term to
remove high frequency noise. We use a fourth-order Runge-Kutta time
integrator for all evolved variables.

\subsubsection{Initial Conditions}

We set up our initial condition from the ADM variables $g_{ij}$,
$K_{ij}$, lapse $\alpha$, and shift $\beta^i$, as provided by the
initial data discussed in Sec.~\ref{sec:initial_models}. From these we
calculate the BSSN quantities via their definition, setting $B^i=0$,
and using cubic extrapolation for $\tilde\Gamma^i$ at the outer
boundary. This extrapolation is necessary since the $\tilde\Gamma^i$ are
calculated from derivatives of the metric, and one cannot use centered
finite differencing stencils near the outer boundary. We assume that
one could instead also use one-sided derivatives to calculate
$\tilde\Gamma^i$ on the boundary.

The extrapolation stencils distinguish between points on the faces,
edges, and corners of the grid. Points on the faces are extrapolated
via stencils perpendicular to that face, while points on the edges and
corners are extrapolated with stencils aligned with the average of the
normals of the adjoining faces. For example, points on the $(+x,+y)$
edge are extrapolated in the $(1,1,0)$ direction, while points in the
$(+x,+y+z)$ corner are extrapolated in the $(1,1,1)$ direction. Since
several layers of boundary points have to be filled for higher order
schemes (e.g., three layers for a fourth order scheme), we proceed
outwards starting from the innermost layer. Each subsequent layer is
then defined via the points in the interior and the previously
calculated layers.

\subsubsection{Boundary Conditions}

During time evolution, we apply a Sommerfeld-type radiative boundary
condition to all components of the evolved BSSN variables as described
in \cite{Alcubierre2000}. The main feature of this boundary condition
is that it assumes approximate spherical symmetry of the solution,
while applying the actual boundary condition on the boundary of a
cubic grid where the face normals are not aligned with the radial
direction. This boundary condition defines the right hand side
of the BSSN state vector on the outer boundary, which is then
integrated in time as well, so that the boundary and interior are
calculated with the same order of accuracy.

The main part of the boundary condition assumes that we have an
outgoing radial wave with some speed $v_0$:
\begin{eqnarray}
  X & = & X_0 + \frac{u(r - v_0 t)}{r}
\end{eqnarray}
where $X$ is any of the tensor components of evolved variables, $X_0$
the value at infinity, and $u$ a spherically symmetric perturbation.
Both $X_0$ and $v_0$ depend on the particular variable and have to be
specified. This implies the following differential equation:
\begin{eqnarray}
  \partial_t X & = & - v^i \partial_i X - v_0\, \frac{X - X_0}{r}\,,
\end{eqnarray}
where $v^i = v_0\, x^i/r$. The spatial derivatives $\partial_i$ are
evaluated using centered finite differencing where possible, and
one-sided finite differencing elsewhere. We use second order stencils
in our implementation.

In addition to this main part, we also account for those parts of the
solution that do not behave as a pure wave, e.g., Coulomb type terms
caused by infall of the coordinate lines. We assume that these parts
decay with a certain power $p$ of the radius. We implement this by
considering the radial derivative of the source term above, and
extrapolating according to this power-law decay.

Given a source term $(\partial_t X)$, we define the corrected source
term $(\partial_t X)^*$ via
\begin{eqnarray}
  (\partial_t X)^* & = & (\partial_t X) + \left( \frac{r}{r - n^i
    \partial_i r} \right)^p\; n^i \partial_i (\partial_t X)\,,
\end{eqnarray}
where $n^i$ is the normal vector of the corresponding boundary face.
The spatial derivatives $\partial_i$ are evaluated by comparing
neighbouring grid points, corresponding to a second-order stencil
evaluated in the middle between the two neighbouring grid points. We
assume a second-order decay, i.e., we choose $p=2$.

As with the initial conditions above, this boundary condition is
evaluated on several layers of grid points, starting from the
innermost layer. Both the extrapolation and radiative boundary
condition algorithms are implemented in the publicly available
\texttt{NewRad} component of the Einstein Toolkit.

This boundary condition is only a coarse approximation of the actual
decay behavior of the BSSN state vector, and it does not capture the
correct behavior of the evolved variables. However, we observe that
this boundary condition leads to stable evolutions if applied
sufficiently far from the source. Errors introduced at the boundary
(both errors in the geometry and constraint violations) propagate
inwards with the speed of light \cite{ES-Brown2007b}. Gauge changes
introduced by the boundary condition, which are physically not
observable, propagate faster, with a speed up to $\sqrt{2}$ for our
gauge conditions.

\subsection{General-Relativistic Hydrodynamics: GRHydro}
\label{sec:hydro}

We employ the open-source GR hydrodynamics code \code{GRHydro} that is
part of the \code{EinsteinToolkit} \cite{einsteintoolkitweb} and is an
updated version of the code \code{Whisky} described in \cite{baiotti:04}.

The equations of ideal GR hydrodynamics evolved by \code{GRHydro} are
derived from the local GR conservation laws of mass and
energy-momentum,
\begin{equation}
  \nabla_{\!\mu} J^\mu = 0, \qquad \nabla_{\!\mu} T^{\mu \nu} = 0\,\,,
  \label{eq:equations_of_motion_gr}
\end{equation}
where $ \nabla_{\!\mu} $ denotes the covariant derivative with respect
to the 4-metric. $ J^{\,\mu} = \rho u^{\,\mu} $ is the mass current
with the 4-velocity $ u^{\,\mu} $ and the rest-mass density $\rho$.  $
T^{\mu \nu} = \rho h u^{\,\mu} u^{\,\nu} + P g^{\,\mu \nu} $ is the
stress-energy tensor.  The quantity $ h = 1 + \epsilon + P / \rho $ is
the specific enthalpy, $P$ is the fluid pressure and $\epsilon$ is the
specific internal energy.

We choose a definition of the 3-velocity that corresponds to the
velocity seen by an Eulerian observer at rest in the current spatial
3-hypersurface \cite{york:83},
\begin{equation}
v^i = \frac{u^i}{W} + \frac{\beta^i}{\alpha}\,\,,
\label{eq:vel}
\end{equation}
where $W = (1-v^i v_i)^{-1/2}$ is the Lorentz factor. In terms of
the 3-velocity, the contravariant 4-velocity is then given by
\begin{equation}
u^0  = \frac{W}{\alpha}\,,\qquad
u^i = W \left( v^i - \frac{\beta^i}{\alpha}\right)\,\,,
\end{equation}
and the covariant 4-velocity is
\begin{equation}
u_0  = W(v^i \beta_i - \alpha)\,,\qquad
u_i = W v_i\,\,.
\end{equation}

The \code{GRHydro} scheme is written in a first-order hyperbolic
flux-conservative evolution system for the conserved variables
$\hat{D}$, $\hat{S}^i$, and $\hat{\tau}$ in terms of the primitive
variables $\rho, \epsilon, v^i$,
\begin{eqnarray}
  \hat{D} &=& \sqrt{\gamma} \rho W,\nonumber\\
  \hat{S}^i &=& \sqrt{\gamma} \rho h W^{\,2} v^i,\nonumber\\
  \hat{\tau} &=& \sqrt{\gamma} \left(\rho h W^{\,2} - P\right) - D\,,
\end{eqnarray}
where $ \gamma $ is the determinant of $\gamma_{ij} $.
The evolution system then becomes
\begin{equation}
  \frac{\partial \mathbf{U}}{\partial t} +
  \frac{\partial \mathbf{F}^{\,i}}{\partial x^{\,i}} =
  \mathbf{S}\,\,,
  \label{eq:conservation_equations_gr}
\end{equation}
with
\begin{eqnarray}
  \mathbf{U} & = & [\hat{D}, \hat{S}_j, \hat{\tau}], \nonumber\\
  \mathbf{F}^{\,i} & = & \alpha
  \left[ \hat{D} \tilde{v}^{\,i}, \hat{S}_j \tilde{v}^{\,i} + \delta^{\,i}_j P,
  \hat{\tau} \tilde{v}^{\,i} + P v^{\,i} \right]\!, \nonumber \\
  \mathbf{S} & = & \alpha
  \bigg[ 0, T^{\mu \nu} \left( \frac{\partial g_{\nu j}}{\partial x^{\,\mu}} - 
  \Gamma^{\,\lambda}_{\mu \nu} g_{\lambda j} \right), \nonumber\\
  & &\qquad\alpha \left( T^{\mu 0}
  \frac{\partial \ln \alpha}{\partial x^{\,\mu}} -
  T^{\mu \nu} \Gamma^{\,0}_{\mu \nu} \right) \bigg]\,.
\end{eqnarray}%
Here, $ \tilde{v}^{\,i} = v^{\,i} - \beta^i / \alpha $ and $
\Gamma^{\,\lambda}_{\mu \nu} $ are the 4-Christoffel symbols.  The
above equations are solved in semi-discrete fashion. The spatial
discretization is performed by means of a high-resolution
shock-capturing (HRSC) scheme employing a second-order accurate
finite-volume discretization. We make use of the Marquina flux formula
for the local Riemann problems and piecewise-parabolic cell interface
reconstruction (PPM\@). For a review of such methods in the GR context,
see~\cite{font:08}. The time integration and coupling with curvature
are carried out with the Method of
Lines~\cite{Hyman-1976-Courant-MOL-report}.

\subsection{Equation of State and Initial Stellar Models}
\label{sec:initial_models}

For the purpose of this study, we employ the simple analytic
\emph{hybrid} EOS \cite{janka:93,dimmelmeier:02}
that combines a 2-piece piecewise polytropic pressure $P_{\rm P}$ with
a thermal component $P_{\rm th}$, i.e., $P=P_{\rm P}+P_{\rm th}$.  To
model the stiffening of the EOS at nuclear density $\rho_{\rm
nuc}\cong2\times10^{14}\,\rm{g}\,\rm{cm}^{-3}$, we assume that the
polytropic index $\gamma$ jumps from $\gamma_1$ below nuclear density
to $\gamma_2$ above.  As detailed in \cite{dimmelmeier:02a}, it is
possible to construct an EOS that is continuous at $\rho_{\rm nuc}$,
\begin{equationarray}
  P & = & \frac{\gamma - \gamma_{\rm th}}{\gamma - 1}
  K \rho_{\rm nuc}^{\gamma_1 - \gamma}
  \rho^{\gamma} - \frac{(\gamma_{\rm th} - 1) (\gamma - \gamma_1)}
  {(\gamma_1 - 1) (\gamma_2 - 1)}
  K \rho_{\rm nuc}^{\gamma_1 - 1} \rho \nonumber \\
  & & + (\gamma_{\rm th} - 1) \rho \epsilon\,.
  \label{eq:hybrid_eos}
\end{equationarray}%
In this, $\epsilon=\epsilon_{\rm P}+\epsilon_{\rm th}$ denotes the
total specific internal energy which consists of a polytropic and a
thermal contribution. $K = 4.897 \times 10^{14}$ [cgs] is the
polytropic constant for a polytrope of relativistic degenerate
electrons at $Y_e=0.5$ . The thermal index $\gamma_{\rm th}=1.5$
corresponds to a mixture of relativistic ($\gamma=4/3$) and
non-relativistic ($\gamma=5/3$) gas.  This EOS mimics the effects of
the stiffening of the physical EOS at $\rho_{\rm nuc}$ and can handle
the significant thermal pressure contribution introduced by shock
heating in the postbounce phase. Provided appropriate choices of EOS
parameters (e.g., \cite{dimmelmeier:07}), the hybrid EOS leads to
qualitatively correct collapse and bounce dynamics. Consequently, this leads to GW
signals that are similar in morphology, characteristic frequencies and
amplitudes to those computed from more microphysically complete
simulations \cite{ott:07prl,dimmelmeier:08, abdikamalov:10}.

We employ $n=3$ ($\gamma_{\rm ini}=\gamma = 4/3$) polytropes in
rotational equilibrium generated via Hachisu's self-consistent field
method \cite{komatsu:89a, komatsu:89b} that not only provides fluid,
but also spacetime curvature initial data. The polytropes are set up
with the rotation law discussed in~\cite{zwerger:97, dimmelmeier:02}
and are parametrized via the differential rotation parameter $ A $ and
the initial ratio $ T/|W| $ of rotational kinetic energy $ T $ to
gravitational binding energy $ |W| $.  While being set up as
marginally stable polytropes with $\gamma_{\rm ini} = 4/3$, during
evolution, the initial sub-nuclear polytropic index $\gamma_{1}$ is
reduced to $\gamma_1 < \gamma_{\rm ini}$ to accelerate
collapse. Following previous
studies~\cite{zwerger:97,dimmelmeier:02,ott:07cqg}, we use $\gamma_2 =
2.5$ in the super-nuclear regime.

From the initial stellar configurations of
\cite{zwerger:97,dimmelmeier:02} we draw a subset of three
models that cover the range of astrophysically expected GW signals
from rotating iron core collapse~\cite{dimmelmeier:08} 
and accretion-induced collapse~\cite{abdikamalov:10}.
Our choices have been used previously in a comparison study
of full GR and conformally-flat simulations~\cite{ott:07cqg}:
\begin{itemize}
\item Model {\bf A1B3G3} is in near uniform rotation with $A=50\times10^3\,\rm{km}$, has $T/|W| =
  0.9\%$, and, once mapped to the evolution grid, uses a sub-nuclear
  adiabatic index $\gamma_1 = 1.31$. Its GW signal is of the standard
  ``Type-I''
  morphology~\cite{zwerger:97,dimmelmeier:02,dimmelmeier:07} and of
  moderate strength (see \cite{dimmelmeier:02,ott:07cqg} for details).

\item Model {\bf A3B3G3} also uses $\gamma_1 = 1.31$. It is strongly
  differentially rotating, with its initial central angular velocity
  dropping by a factor of two over $A=500\,\mathrm{km}$. This, in
  combination with $T/|W| = 0.9\%$, leads to rapid rotation in the
  inner core, resulting in a very strong GW signal at core bounce and
  dynamics that are significantly affected by centrifugal effects. It
  produces a ``Type-I'' GW signal with a centrifugally-widened broad
  peak at core bounce.

\item Model {\bf A1B3G5} has the same rotational setup as model
  A1B3G3, but its sub-nuclear adiabatic index is reduced to $\gamma_1
  = 1.28$. This leads to rapid collapse, to a very small inner core at
  core bounce, and to a weak ``Type-III'' GW signal
  \cite{zwerger:97,dimmelmeier:02} akin to that potentially emitted by an
  accretion-induced collapse event~\cite{abdikamalov:10}.

\end{itemize}
For convenience, key model parameters are summarized in
Table~\ref{table:models}.
\begin{table}[t]
\caption{Initial parameters of differentially rotating stellar cores used
for the core collapse simulations. The models are described by 
three quantities: the degree of differential rotation $A$, 
the ratio $T/|W|$ of rotational to potential energy,
and the sub-nuclear adiabatic index $\gamma_1$ during the collapse. For convenience we also report 
the wave-signature type of the three models and the mass $M$ present on the computational grid.}
\label{table:models}
\begin{ruledtabular}
\begin{tabular}{llcccc}
Model & Type &$A\, [10^3\, \rm{km}]$ & $T/|W|\, [\%]$ & $\gamma_1$ & $M\,[M_\odot]$ \\
\hline
A1B3G3 & I (weak)   & $50.0$ & $0.89$ & $1.31$ & $1.46$ \\
A1B3G5 & III        & $50.0$ & $0.89$ & $1.28$ & $1.46$ \\
A3B3G3 & I (strong) &  $0.5$ & $0.89$ & $1.31$ & $1.46$ \\
\end{tabular}
\end{ruledtabular}
\end{table}

\section{Gravitational Wave Extraction}
\label{sec:waveextract}

\subsection{The Quadrupole Approximation}
\label{sec:quad}

The quadrupole approximation is the only means of extracting
GWs in Newtonian or conformally-flat GR simulations, but has found
wide application also in GR simulations of stellar collapse
\cite{shibata:04,shibata:05,ott:07cqg,ott:07prl}.

The coordinate-dependent quadrupole formula estimates the GW strain
seen by an asymptotic observer by considering exclusively the
quadrupole stress-energy source. It neglects any non-linear GR
effects. This approximation is valid strictly only in the weak-field
$\frac{G}{c^2}\frac{R}{M}\ll1$ and slow-motion $\frac{v}{c}\ll1$ limit
\cite{thorne:80} where spacetime is essentially flat.

The quadrupole formula is given in the transverse-traceless (TT) gauge
by
\begin{equation} \label{eq:quad-strain}
  h_{jk}^{TT}(t,\mathbf{x})=\frac{2}{c^4}\frac{G}{R}\left[\frac{d^2}{dt^2}I_{jk}(t-R/c)
  \right]^{TT}\,,
\end{equation}
where 
\begin{equation} \label{eq:quad-mass}
  I_{jk}=\int\tilde{\rho}(t,\mathbf{x})\left[x_j x_k -
    \frac{1}{3}x^2\delta_{jk}\right]d^3x
\end{equation}
denotes the reduced mass-quadrupole tensor.  Since we are working in
the weak-field, slow-motion approximation, the placement of tensor
indices is arbitrary.  $I_{jk}$ is not uniquely defined in GR and the
choice of the density variable $\tilde{\rho}$ is not obvious.
Following
\cite{dimmelmeier:02,shibata:04,dimmelmeier:08,ott:07cqg,ott:07prl},
we set $\tilde{\rho} = \sqrt{\gamma}\, W \rho = \hat{D}$, because, (i),
this is the conserved density variable in our code, and (ii),
$\sqrt{\gamma}\, d^3 x$ is the natural volume element. See
\cite{baiotti:09} for other potential choices and their relative
performance for GWs from oscillating polytropes.

The reduced mass-quadrupole tensor can be computed directly from the
computed distribution $\hat{D}(t,\mathbf{x})$.  In order to eliminate
the effects of numerical noise when differentiating
Eq.~(\ref{eq:quad-mass}) twice in time, we make use of the continuity
equation to obtain the first time derivative of
Eq.~(\ref{eq:quad-mass}) without numerical differentiation
\cite{finnevans:90,blanchet:90,dimmelmeier:05},
\begin{equation}
  \frac{d}{dt}I_{jk}=\int \hat{D}(t,\mathbf{x})\left[\tilde{v}^j x^k + \tilde{v}^k x^j - \frac{2}{3}(x^l\tilde{v}^l)\delta^{jk}\right]d^3x\,\,,
\label{eq:dtquad}
\end{equation}
where we set $\tilde{v}^i = v^i$ as defined by Eq.~\ref{eq:vel}.  Note
that we have switched to contravariant variables in the integrand as
these are the ones present in the code.  Finally, the remaining time
derivative needed for evaluating the quadrupole GW strain
(Eq.~\ref{eq:quad-strain}) is performed numerically.

In order to assess the sensitivity of the predicted waves on the
particular choice of the velocity variable $\tilde{v}^i$ in
Eq.~(\ref{eq:dtquad}), we implement two modified versions.  In variant
\emph{VS}, we use Shibata~et~al.'s definition of the 3-velocity (e.g.,
\cite{shibata:04}) that differs from ours by a gauge term. In variant
\emph{PV}, we follow \cite{dimmelmeier:02a} and employ physical
velocity components (individually bound to $v < c$) that, in Cartesian
coordinates, are given by $\{v_x, v_y, v_z\} \approx
\{\sqrt{\gamma_{11}} v^1, \sqrt{\gamma_{22}} v^2, \sqrt{\gamma_{33}}
v^3\}$, assuming that the 3-metric is nearly diagonal (which is the
case in our gauge; see Sec.~\ref{sec:curv}).

\subsection{The Regge-Wheeler-Zerilli-Moncrief Formalism}

A particular \textit{Ansatz} for analyzing gravitational radiation in
terms of odd and even multipoles in the far-field of the
source was originally developed by Regge, Wheeler
\cite{Regge1957} and Zerilli \cite{Zerilli1970b, Zerilli:1971wd},
respectively. Moncrief subsequently provided a gauge-invariant
reformulation \cite{Moncrief1974} (see \cite{nagar:05} for a review).
Assuming that, at large distances from the source, the GW content of the spacetime can be
viewed as a linear perturbation to a fixed background, we can write
\begin{equation}
g_{\mu\nu}=g^0_{\mu\nu}+h_{\mu\nu}\,,
\end{equation}
where $g^0_{\mu\nu}$ is the fixed background metric
and $h_{\mu\nu}$ its linear perturbation.
The background metric $g^0_{\mu\nu}$ is
usually assumed to be of Minkowski or Schwarzschild form, which we can write as
\begin{equation}
ds^2=-N dt^2 + A dr^2 + r^2(d\theta^2 + \sin^2\theta d\phi^2)\,.
\end{equation}

By splitting the spacetime into timelike and radial
and angular parts, it is possible to decompose
the metric perturbation $h_{\mu\nu}$ into odd
and even multipoles, i.e.,~we can write
\begin{equation}
h_{\mu\nu}=\sum_{\ell m}\left[\left(h_{\mu\nu}^{\ell m}\right)^{(o)} + \left(h_{\mu\nu}^{\ell m}\right)^{(e)}\right]\ .
\end{equation}
The even and odd multipole components are defined according to their behavior
under a parity transformation $(\theta,\phi)\rightarrow(\pi-\theta,\pi+\phi)$.
Odd multipoles transform as $(-1)^{\ell+1}$ while even multipoles transform as $(-1)^\ell$.
Both multipole components can be expanded
in terms of vector and tensor spherical harmonics (e.g., \cite{thorne:80}).

Given the Hamiltonian of the perturbed Einstein equations in ADM form \cite{Arnowitt62}, 
it is then possible to derive variational principles for the odd and 
even-parity perturbations \cite{Moncrief1974} to 
give equations of motions that are similar to wave equations with a scattering potential.

The solutions to the odd and even-parity wave equations are given by
the Regge-Wheeler-Moncrief and the Zerilli-Moncrief master functions, respectively.
The odd-parity Regge-Wheeler-Moncrief function reads
\begin{eqnarray} \label{eq:Qodd}
Q^{\times}_{\ell m} &\equiv& \sqrt{\frac{2(\ell+1)!}{(\ell-2)!}}
	\frac{1}{r}\left(1-\frac{2M}{r}\right) \nonumber \\
	& & \left[(h_{1}^{\ell m})^{({\rm o})}+\frac{r^2}{2} \partial_r
	\left(\frac{(h_2^{\ell m})^{({\rm o})}}{r^2}\right)\right]\ ,
\end{eqnarray}
and the even-parity Zerilli-Moncrief function reads
\begin{eqnarray} \label{eq:Qeven}
{Q}^{+}_{\ell m} \equiv \sqrt{\frac{2(\ell+1)!}{(\ell-2)!}}
\frac{r q_1^{\ell m}}{\Lambda\left[r\left(\Lambda-2\right)+6M\right]} \,,
\end{eqnarray}
where $\Lambda=\ell(\ell+1)$, and where
\begin{equation}
\label{q1}
q_1^{\ell m}  \equiv r\Lambda\kappa_1^{\ell m} + \frac{4r}{A^2}\kappa_2^{\ell m} \,,
\end{equation}
with
\begin{eqnarray}
\label{kappa1}
\kappa_1^{\ell m} & \equiv & K^{\ell m}+\frac{1}{A}\left(r\partial_r G^{\ell m}-
	\frac{2}{r}(h_1^{\ell m})^{({\rm e})}\right)\ ,\\
\label{kappa2}
\kappa_2^{\ell m} & \equiv &\frac{1}{2}\left[A H_2^{\ell m}-
	\sqrt{A} \partial_r \left(r \sqrt{A} K^{\ell m}\right)\right]\, .
\label{def:q1}
\end{eqnarray}

These master functions depend entirely on the spherical part of the metric given by the coefficients $N$ and
$A$, and
the perturbation coefficients for the
individual metric perturbation components $(h_{1}^{\ell m})^{({\rm o})}$, $(h_{2}^{\ell m})^{({\rm o})}$,
$(h_{1}^{\ell m})^{({\rm e})}$, $(h_{2}^{\ell m})^{({\rm e})}$, 
$H_0^{\ell m}$, $H_1^{\ell m}$, $H_2^{\ell m}$, $K^{\ell m}$, and $G^{\ell m}$ which
can be obtained
from any numerical spacetime by projecting out the Schwarzschild or Minkowski
background \cite{Camarda:1998wf}.
For example, the coefficient $H_2^{\ell m}$ can be obtained via
\begin{equation}
H_2^{\ell m}=\frac{1}{A}\int (g_{rr}-A) Y_{\ell m}\,d\Omega\,,
\end{equation}
where $g_{rr}$ is the radial component of the numerical metric represented in the spherical-polar coordinate basis,
$Y_{\ell m}$ are spherical harmonics, and $d\Omega$ is the surface line element of the $S^2$ extraction sphere.
The coefficient $A$ represents the spherical part of the background metric
and can be obtained by projection of the numerical metric component $g_{rr}$ on $Y_{00}$ over the extraction sphere
\begin{equation} \label{eq:H2lm}
A=\frac{1}{4\pi}\int g_{rr} d\Omega\,.
\end{equation}
Similar expressions hold for the remaining perturbation coefficients.

The odd- and even-parity master functions Eq.~(\ref{eq:Qodd}) and Eq.~(\ref{eq:Qeven}) can be
straight-forwardly related to the gravitational-wave strain and are given by
\begin{eqnarray}
\label{eq:h-Q}
h_{+}-\mathrm{i}h_{\times}&=&\frac{1}{\sqrt{2}r}\sum_{\ell,m}\left(
	Q^{+}_{{\ell m}} - \mathrm{i}\int_{-\infty}^{t}
	Q^{\times}_{{\ell m}}(t')dt'\right) \nonumber \\
	& & \;_{_{-2}}Y^{{\ell m}}(\theta,\phi)
	+ {\cal O}\left(\frac{1}{r^2}\right)\, ,
\end{eqnarray}
where ${_{-2}}Y^{{\ell m}}(\theta,\phi)$ are the spin-weight $s=-2$
spherical harmonics.  We note that this relation is strictly true only
at an infinite distance from the source.  Since our numerical domain is
finite in size, we choose some, ideally large, but finite radius.  In
Sec.~\ref{sec:radial}, we check how well the GWs extracted with the
RWZM formalism asymptote with increasing extraction radius.

In the present work, our models exclusively trigger the even-parity master function $Q^+$, and
$Q^\times$ is zero. In this case, we can 
simplify Eq.~(\ref{eq:h-Q}) and obtain
\begin{equation}
\label{eq:h-Q-simple}
h_{+,e} =\frac{1}{\sqrt{2}r}
	Q^{+}_{{2 0}} \;_{_{-2}}Y^{{2 0}}(\theta,\phi)\, ,
\end{equation}
relating the strain directly to $Q^+$.

\subsection{Newman-Penrose Scalars}

Another method for calculating the gravitational waveforms is based on
the conformal structure of asymptotically flat spacetimes
as established by Bondi, Sachs and Penrose \cite{Bondi1962, Sachs1962, Penrose1963}.
This method is conveniently represented in terms of spin-weighted scalars
as introduced by Newman and Penrose \cite{Newman1962}. In the following we refer to it as NP extraction.
According to the peeling theorem \cite{Bondi1962, Sachs1962}, 
a certain component of the conformal Weyl tensor obeys the slowest $1/r$
fall-off from the source, and hence is identified as
outgoing gravitational radiation:
\begin{equation}
C_{\alpha\beta\gamma\delta}=\frac{\Psi_4}{r}+\frac{\Psi_3}{r^2}+\frac{\Psi_2}{r^3}+\frac{\Psi_1}{r^4}+\frac{\Psi_0}{r^5}+\mathcal{O}(r^{-6})\,.
\end{equation}
Here, the slowest fall-off is obeyed by the
NP scalar $\Psi_4$, which is defined as\footnote{Our 
definition proceeds along the lines
of Appendix C of Ref.~\cite{Sperhake2006} but for comparison
with the quadrupole results, we define the
Newman-Penrose scalar with the opposite sign of their Eq.~(C1).}
\begin{equation}
  \Psi_4 \equiv -C_{\alpha \beta \gamma \delta} n^{\alpha}
      \bar{m}^{\beta} n^{\gamma} \bar{m}^{\delta},
\end{equation}
where $C_{\alpha \beta \gamma \delta}$ is the conformal Weyl tensor
associated with the 4-metric $g_{\alpha \beta}$ and $n$, $\bar{m}$ are
part of a null-tetrad \cite{Penrose1963, Newman1962}
$\ell,~n,~m,~\bar{m}$ which satisfies $-\ell\cdot n=1 =m\cdot \bar{m}$
while all other inner products vanish. In addition, this tetrad is
related to the 4-metric via
$g_{ab}=l_{a}n_{b}+l_{b}n_{a}-m_{a}\bar{m}_{b}-m_{b}\bar{m}_{a}$.  At
future null infinity $\scri^+$, the topology of the spacetime is a
time succession of spheres, $S^2\times\mathbb{R}$. Hence the simplest
choice for the null tetrad at $\scri^+$ is such that it resembles the
unit sphere metric. Moreover, the simplest choice for a coordinate
system at $\scri^+$ is given by the Bondi gauge \cite{Bondi1962,
Sachs1962}, which makes use of an areal radius coordinate.

In most current numerical relativity simulations, the
radiation is computed at a \textit{finite} radius where the Bondi
coordinates are usually not imposed. Rather, we use the gauge as
evolved by the $1+\log$ slicing and $\Gamma$-driver conditions
discussed in Sec.~\ref{sec:curv}. In practice, we impose a simple
polar-spherical coordinate system with constant coordinate radius
$R^2=x^2+y^2+z^2$, which does not take into account the background
geometry, and hence does not make use of an areal radius.  Thus, the
gravitational radiation as computed on these coordinate spheres is not
measured in the correct gauge, and leads to a systematic error that
needs to be assessed.  Note that it is principally possible to
transform to the correct gauge \cite{Lehner2007}.

In our construction of an approximate tetrad, we follow common
practice (e.g.~\cite{Pollney:2009yz,Brugmann:2008zz,Baker:2001sf}) and
use a triad of spatial vectors $u,~v,~w$ obtained via a Gram-Schmidt
orthonormalization starting from
\begin{eqnarray}
  u^i &=& [x,~y,~z], \\
  v^i &=& [xz,~yz,~-x^2-y^2], \\
  w^i &=& \epsilon^i{}_{mn} v^m w^n,
\end{eqnarray}
where $x,~y,~z$ are Cartesian coordinates of the computational grid and
$\epsilon^{imn}$ is the three-dimensional Levi-Civita symbol. The tetrad
is given in terms of this triad and the timelike normal
vector $\hat{n}^{\alpha}$ by
\begin{eqnarray}
  \ell^{\alpha} &=& \frac{1}{\sqrt{2}}(\hat{n}^{\alpha}+u^{\alpha}), \\
  n^{\alpha} &=& \frac{1}{\sqrt{2}}(\hat{n}^{\alpha} -u^{\alpha}), \\
  m^{\alpha} &=& \frac{1}{\sqrt{2}}(v^{\alpha} + iw^{\alpha}).
\end{eqnarray}

A straightforward calculation shows that we are thus able to
express $\Psi_4$ exclusively in terms of the ``3+1'' variables according to
\begin{eqnarray}
  &&\Psi_4 = \frac{1}{2} \left[ E_{mn}(w^m w^n - v^m v^n) + B_{mn}
      (v^m w^n + w^m v^n) \right] \nonumber \\
  && + \frac{i}{2}
      \left[E_{mn}(v^m w^n - w^m v^n) + B_{mn} (w^m w^n + v^m v^n)\right],
  \label{eq:Psi4}
\end{eqnarray}
where the electric and magnetic parts of the Weyl tensor are defined as
\cite{Friedrich1996}
\begin{eqnarray}
  E_{\alpha \beta} &\equiv& \bot^{\mu}{}_{\alpha} \bot^{\nu}{}_{\beta}
      C_{\mu \rho \nu \sigma} \hat{n}^{\rho} \hat{n}^{\sigma}, \\
  B_{\alpha \beta} &\equiv& \bot^{\mu}{}_{\alpha} \bot^{\nu}{}_{\beta}
      {}^*C_{\mu \rho \nu \sigma}.
\end{eqnarray}
Here the ${}^*$ denotes the Hodge dual and $\bot^{\mu}{}_{\alpha}
\equiv \delta^{\mu}{}_{\alpha} + \hat{n}^{\mu}\hat{n}_{\alpha}$
is the projection operator. The Gauss-Codazzi equations (see e.g.~\cite{Alcubierre:2008}) enable us
to calculate the electric and magnetic parts from the ``3+1'' variables
according to
\begin{eqnarray}
  E_{ij} &=& R_{ij} - \gamma^{mn} (K_{ij} K_{mn} - K_{im}K_{jn}), \\
  B_{ij} &=& \gamma_{ik} \epsilon^{kmn} D_m K_{nj}.
\end{eqnarray}

In a given numerical simulation, we calculate $\Psi_4$ from Eq.~(\ref{eq:Psi4})
on a set of coordinate spheres defined by $R_{\rm ex}={\rm const}$. On
each of these spheres, we use spin-weighted spherical harmonics ${}_{-2}Y_{\ell m}(\theta,\phi)$ of
spin weight $-2$ in order to decompose the resulting wave signal
into multipoles
\begin{eqnarray}
  \Psi_4(t,~\theta,~\phi) = \sum_{\ell, m} \Psi^{\ell m}_4(t) {}_{-2}Y_{\ell m}(\theta, \phi),
      \nonumber \\
  \Psi^{\ell m}_4(t) = \int \Psi_4(t,\theta,\phi) \bar{{}_{-2}Y_{\ell m}}(\theta,\phi)
      d\Omega.
  \label{eq:Psi4multipoles}
\end{eqnarray}
In all our simulations, the wave signal is dominated by the $\ell=2$, $m=0$
mode whose angular dependence is given by
\begin{equation}
  {}_{-2}Y_{20} = \sqrt{\frac{15}{32\pi}} \sin^2 \theta.
\end{equation}

The NP scalar $\Psi_4$ is related to the gravitational
wave strain via
\begin{equation} \label{eq:psi4-strain}
  \Psi_4 = \ddot{h}_+ - i\ddot{h}_{\times}.
\end{equation}
It is convenient to decompose the two GW polarizations into multipoles
in analogy to Eq.~(\ref{eq:Psi4multipoles})
\begin{equation}
  h_+ - ih_{\times} = \sum_{\ell=2}^{\infty} \sum_{m=-\ell}^\ell
      \left[ h_+^{(\ell m)}(t) - ih_{\times}^{(\ell m)}(t) \right]
      {}_{-2}Y_{\ell m}(\theta,\phi).
\end{equation}
These multipoles are related to those of the NP
scalar by
\begin{equation}
  \Psi^{\ell m}_4 = \ddot{h}_+^{(\ell m)} - i\ddot{h}_{\times}^{(\ell m)}.
\end{equation}

Note that the final result is not fully gauge-invariant
and contains an unknown amount of systematic error.
The reasons are two-fold: First, we did not choose
a proper Bondi null tetrad on our extraction spheres,
and, second, the extraction spheres have finite radius,
thus are neglecting non-linear back-scattering effects
of the gravitational field in the wave zone.
However, since our coordinate frame asymptotically approaches the Minkowski spacetime,
both errors can be minimized by
performing extrapolation to future null infinity $\scri^+$,
using a set of extraction spheres at finite radii. 
Unfortunately, even if the extrapolation is accurate,
an uncertain amount of residual error may remain.
In Sec.~\ref{sec:radial}, we check how well the extracted waves 
approximate their asymptotic shape and magnitude.

\subsection{Cauchy-Characteristic Extraction}

To circumvent the problem of finite-radius extraction and to eliminate
this systematic error, we apply the technique of CCE \cite{Bishop98b,
  Babiuc:2005pg, Winicour05, Reisswig:2009us, Reisswig:2009rx,
  Babiuc:2010ze} to obtain the NP scalar $\Psi_4$ as discussed in the
previous section\footnote{ Alternatively during this procedure, we
  also compute the \textit{Bondi news function} \cite{Newman1962}
  $\news$, which is related to the GW strain by only one integration
  in time. We find that the news function is less robust
  when residual matter is present at the world-tube location and
  therefore restrict our attention to $\Psi_4$ only.}, in
this case directly evaluated at future null infinity $\scri^+$.  The
CCE technique couples an exterior characteristic evolution of the full
Einstein equations to the interior strong-field 3+1 Cauchy evolution
of the spacetime.

Characteristic evolutions are based on null-hypersurface foliations of
spacetime and have the advantage of allowing for a compactification of
the radial coordinate component, thus allowing to include future null
infinity $\scri^+$ on the computational grid \cite{Winicour05}.
Unfortunately, the characteristic formulation gives rise to the
formation of caustics, i.e., the null rays on which the coordinate
system is based can intersect in strong-field regions, leading to
coordinate singularities.  The scheme is therefore not well suited for
the evolution of the actual GW source.  Characteristic evolutions, on
the other hand, are well adapted to the far-field region of spacetime
and can efficiently evolve the metric fields out to $\scri^+$ where it
is possible to obtain $\Psi_4$ (and, hence, $h$) in a mathematically
unambiguous and gauge-invariant way \cite{Bishop97b, Babiuc:2005pg,
  Babiuc:2008qy, Bishop03}.

We therefore proceed as follows: we evolve the interior region
containing the collapsing matter with the standard 
Cauchy formulation as described in Sec.~\ref{sec:curv} and \ref{sec:hydro}.
During the Cauchy evolution, we store the 3-metric components
including lapse and shift on coordinate spheres with fixed 
radius $R_\Gamma$ defining the world-tube $\Gamma$.

This world-tube forms the inner boundary for the subsequent
characteristic evolution of the Einstein equations. The full 4-metric
can be reconstructed from the stored 3-metric components, the lapse and
the shift at the inner boundary. Upon construction of proper initial data
on an initial null hypersurface, which here we simply assume to be
conformally flat, we have then fully specified any data necessary to
evolve the fields out to $\scri^+$.  More details on the exact
mathematical procedure can be found in \cite{Bishop98b, Winicour05}.
The characteristic field equations are solved numerically using the
{\tt PITT} null evolution code \cite{Bishop97b}.  The numerical
implementation of CCE including results from binary black hole mergers
is discussed in \cite{Reisswig:2009us, Reisswig:2009rx, Babiuc:2010ze}.
For the characteristic computational grid, we use $N_r=397$ points in the radial direction
and $N_{\rm ang}=73$ points in each angular direction for the two stereographic
patches covering the sphere. The characteristic timestep size equals that of the Cauchy evolution.

After each characteristic timestep, the NP scalar $\Psi_4$ is evaluated directly from the metric at
$\scri^+$ and transformed to the desired Bondi gauge \cite{Babiuc:2008qy}.  Thus, the CCE
method is free of gauge and near-zone effects and represents the
most rigorous extraction technique.  However, there is still some
remaining systematic error that is due to the presence of matter at the world-tube locations. 
Since the current set of
characteristic equations does not take into account any form of matter
contribution, a non-zero stress-energy tensor introduces
an unknown error.  
We therefore have to perform checks of the dependence
of the waveforms on the world-tube locations.
In principle, it is possible to also incorporate matter
on the characteristic side \cite{Bishop99}, which we
leave to future work.

We note that CCE does not remove the artifical outer grid
boundary from the Cauchy evolution. Thus, inconsistencies arising from
this boundary can, in principle, still influence the interior domain.
It is possible to circumvent this problem by enlarging the
computational domain so that the outer boundary is causally
disconnected from the world-tube locations (see,
e.g., \cite{Pollney:2009yz} in the context of binary black holes).
In simulations of core collapse, however, this is currently 
not computationally feasible, but experiments with varied outer boundary
locations have shown that boundary effects are negligible for our
current choice of boundary location.

Finally, we point out that inconsistencies in the characteristic and
Cauchy initial data may lead to a loss of some non-linear
effects. Even though we expect these problems to be very small (see
\cite{Reisswig:2009rx}). These and the outer boundary issues
highlighted in the above can be fully accounted for only by
employing Cauchy-characteristic matching (CCM)
(e.g. \cite{Winicour05}).  This technique uses the characteristic
evolution as a generator for Cauchy boundary data at the world-tube,
i.e.~the world-tube becomes a two-way boundary between Cauchy and
characteristic evolution.  In practice, CCM has not yet been
successfully implemented.

\subsection{Remarks on Integration and Physical Units}
\label{sec:int-units}

The NP scalar $\Psi^{\ell m}_4$ must be integrated twice in time to
yield the strain $h$, which introduces an artificial ``memory''
\cite{Berti2007}, i.e., a non-linear drift of the signal so that the
wavetrain deviates from an oscillation about zero. This behavior
cannot be explained by the two unknown integration constants
resulting, at most, in a linear drift.

As suggested in \cite{Reisswig:2010di}, this non-linear drift is a
consequence of random-walk-like behavior induced by numerical
noise. In the present work, we make use of methods that are
strictly valid only in pure vacuum and at our extraction spheres
the average matter densities are non-zero (see
Table~\ref{tab:rho}).  This systematic error can lead to an additional artificial
low-frequency drift.  In order to eliminate this effect, we
use \textit{fixed-frequency integration} (FFI) as proposed in
\cite{Reisswig:2010di}. The NP variable $\Psi^{20}_4(t)$ is Fourier
transformed, the resulting spectrum is divided by $f_{0}^2$ for
frequencies $f<f_{0}$ and divided by $f^2$ otherwise.  An inverse
Fourier transform then yields the strain $h_{\ell m}$ 
essentially free of spurious drifts
and oscillations, given a proper choice of $f_0$.

Finally, we need to address the question of units. The gravitational
wave strain $h_+ -ih_{\times}$ is by construction dimensionless.  For
comparison of waveforms at different extraction radii, it is
convenient to compensate for the $1/D$ fall-off of the
strain, where $D$ is the distance from the observer to the source, and to
work with $Dh_+^{(\ell m)}$ and $Dh_{\times}^{(\ell m)}$ instead.  In most of the
following, we convert from code units, which are in $c=G=M_\odot = 1$,
to cgs units when stating and plotting numerical results. The
conversion factors we use are $1~M_{\odot}=1.4772\times 10^5~{\rm cm}$
for length, and $1~M_{\odot}=0.004927\,\mathrm{ms}$ for time. 
For simplicity, we state the radii of GW extraction spheres and world-tube radii
in code units. These and their corresponding cgs values are listed
in Table~\ref{tab:rho}.


\section{Results}
\label{sec:results}

In this section, we compare the most reliable extraction method that contains the least
amount of systematic errors, CCE,
with the various other
curvature-based extraction methods, i.e.,~RWZM and NP
extraction (both at finite radii). We also
perform a comparison with the quadrupole approximation which has been employed in
virtually all core collapse simulations to date.

This section is structured as follows.  
First, we review briefly the morphology of the gravitational waveforms 
expected from rotating core collapse and bounce.

Second, we elaborate in more
detail on the method with which we obtain the gravitational strain $h$
from the quantities measured during the simulation. This is important,
since the derived strain typically contains severe non-linear drifts making a
proper analysis largely impossible without significant preprocessing.

Third, we assess the accuracy of each individual method, i.e., we
analyze the radial dependence of the NP and RWZM
extraction methods, since they are strictly valid only in an asymptotic frame
at an infinite distance from the source where any contributions from the
stress-energy tensor vanish. Since the matter densities are non-zero at the CCE world-tube
locations, we also analyze the radial dependence of the waveforms extracted via CCE\@.

Fourth, we compare the results obtained via NP and RWZM extraction,
and the approximate QF with results obtained via CCE.  

Finally, we perform a convergence check on the computed waveforms by
using a set of three different resolutions.

\subsection{Morphology of Rotating Core Collapse Waveforms}

The core collapse models considered in this work remain nearly axisymmetric
during collapse and emit GWs predominantly via the even-parity
$(\ell,m)=(2,0)$ spherical harmonic mode.  This mode has a maximum on
the equator and, hence, we plot all waveforms as seen by an observer
in the equatorial plane. We write $h_{+,e}\equiv
h_+^{20}{}_{-2}Y_{20}(\theta=\pi/2,\phi)$ where ${}_{-2}Y_{20}$ is the
spin-weighted spherical harmonic with spin $s=-2$.  Note that the
$(\ell,m)=(2,0)$ mode is axisymmetric and thus, the equatorial strain
$h_{+,e}$ has no $\phi$-dependence.

We convert to cgs units by using the transformation as discussed in
Sec.~\ref{sec:int-units}, and we align the maxima of the waveforms
such that they occur at $t=0.0\,\rm{ms}$, corresponding roughly to the
time of core bounce in each model.

The waveforms of the three models are shown 
in Figs.~\ref{fig:radial_dep_A1B3G3} (model A1B3G3),~\ref{fig:radial_dep_A1B3G5} (model A1B3G5),
and~\ref{fig:radial_dep_A3B3G3} (model A3B3G3). All models exhibit a very similar behavior.
Prior to core bounce $(t<0\, \rm{ms})$, the matter undergoes an aspherical accelerated
collapse. Due to this aspherical acceleration, the GW signal
is monotonically rising until it peaks when the contracting inner core is drastically decelerated.
This deceleration is caused by the sudden stiffening of the EOS 
as a result of nuclear repulsive forces which emerge 
when nuclear densities are reached. 
During this deceleration, the GW
signal becomes rapidly negative, reaching its second peak (the ``bounce peak'') roughly when the core rebounds.
Subsequently, the inner core undergoes a relaxation phase (ring-down) in which it loses its remaining
pulsation energy by launching secondary shocks. This results in an oscillatory GW signal
that decreases in amplitude as the core approaches its final equilibrium.

While the overall morphology of the GWs emitted by the three models is the same,
there are subtle differences that are worth commenting on. Models A1B3G3 and A3B3G3 produce so-called
type-I signals \cite{zwerger:97} with a single pronounced major peak at core bounce.
Since model A3B3G3 is more rapidly spinning, its inner core is more deformed, and hence produces a stronger
GW signal at core bounce than model A1B3G3.
Model A1B3G5 has a very small inner core at bounce and
produces a type-III signal that is characterized by a much less pronounced
bounce peak and generally low-amplitude GW emission.
Note that type-II signals, characterized by \emph{multiple} wide and pronounced bounce peaks
seen in early work \cite{zwerger:97, moenchmeyer:91, kotake:03, yamada:99, ott:04}
have been demonstrated to disappear in simulations using general relativity and a proper
electron-capture treatment \cite{ott:07prl,dimmelmeier:08}.

\subsection{Computing the Strain}

We first consider the computation of the strain $h_{+,e}$ from the
RWZM formalism.  Since our models emit GWs predominantly in the
$(\ell,m)=(2,0)$ even mode $Q^+$, the computation of the strain from
the even- and odd-parity RWZM master functions reduces to
Eq.~(\ref{eq:h-Q-simple}) so that no time integral is necessary to
obtain $h_{+,e}$.  However, we still notice an unphysical drift in the
waves. Since the RWZM master functions are computed at a
\textit{finite} distance from the source, we have the following
systematic errors (summarized as the ``finite-radius error''): (i) a
non-vanishing matter density at the extraction spheres, (ii) near-zone
effects and (iii) gauge ambiguities. The latter error arises
as a result of deviations from the Bondi gauge (see \cite{Lehner2007} 
for an improvement). 
The artificial drift is part of
the finite-radius error, since it is becoming less pronounced with
increasing extraction radius.

In order to reduce the contribution of these artificial low-frequency
components, we first transform to the Fourier domain, multiply by $f$
in order to take the first time derivative, and then apply fixed
FFI \cite{Reisswig:2010di} to obtain
$h_{+,e}$.  This procedure effectively acts as a filter that suppresses
unwanted low-frequency components and at the same time minimizes
spurious oscillations in the time domain such as Gibbs ringing or
additional non-linear low-frequency drifts.

The QF (cf.~Eq.~\ref{eq:quad-strain}) directly computes the strain and
does not suffer from low-frequency drifts. However, the NP and CCE
methods compute the second time derivative of the strain and, hence,
must be integrated twice in time (Eq.~\ref{eq:psi4-strain}). For this,
we employ FFI to minimize the influence of artificial low-frequency
components.  Unfortunately, the time integration is still subject to some
amount of low-frequency error as we shall discuss in the following.

\subsubsection{Error in the Time Integration}
\label{sec:int}

FFI introduces a free parameter $f_0$ that must be chosen based on the
expected lowest \textit{physical} frequency component of the
signal. It must be larger than the spurious artificial low-frequency
contributions that are introduced by aliased numerical noise and
spectral leakage \cite{Reisswig:2010di}.

Unfortunately, in the case considered here, the artificial
low-frequency contributions overlap with the low-frequency part of the
physical signal.  Since it is not possible to disentangle physical
from artificial contributions at a given frequency, we have to choose
an $f_0$ that is larger than the highest \emph{unphysical} frequency
contained in the signal.  Thus, part of the overlapping
\emph{physical} low-frequency spectrum is lost when constructing the
strain $h$.

We identify the highest unphysical frequency by choosing a set of
different $f_0$ for a given waveform, i.e.,~we introduce a family of
strains $h(t;\,f_0)$, and by imposing a relative maximum deviation $\max_t\delta
h(t;\,f_0)/\delta f_0$ between the resulting strains $h(t;\,f_0)$
during ring-down that is not larger than some small $\epsilon$. Since
the ring-down phase is at the end of the wave train, the impact of the
accumulated drift is largest here and can be clearly identified.
Increasing $f_0$ reduces the drift in the ring-down phase, but also
removes physical content, i.e.,~the monotonic rise of the signal in
the prebounce phase.  In order to gauge how much information is lost
prior to core bounce, we compute the differences of $h(t;\,f_0)$ from
the quadrupole waveform in an interval $t\in[-10\,\rm{ms},\,
0\,\rm{ms}]$, since the quadrupole waveforms do not suffer from time
integration issues and are presumably accurate up to at least
the late prebounce phase.  If we stop at some level of tolerance for any
deviations $\max_t\delta h(t;\,f_0)/\delta f_0$ during ring-down and
deviations from the quadrupole waveform prior to core bounce, we have identified an
appropriate $f_0$.  In practice, we choose a threshold
$\max_t\delta Dh(t;\,f_0)/\delta f_0 \lesssim \epsilon\sim0.1\, \rm{cm}/\rm{Hz}$ during ring-down.

Our numerical experiments show that the cut-off
frequency $f_0$ is model and extraction-method dependent and must be determined individually
for any new set of initial data. In Table~\ref{tab:f0}, we list the
frequencies $f_0$ for each of the considered models and extraction
methods which yield the lowest deviations during ring-down and at the
same time resemble as closely as possible the quadrupole waveform in
the prebounce phase.

\begin{table}[t]
\caption{Lowest possible physical frequencies that result in strain
amplitudes with deviations $\max_{t} |\delta Dh(t;\,f_0)/\delta f_0|$
of no more than $\epsilon\sim0.1\, \rm{cm}/\rm{Hz}$.  In all cases,
CCE yields the lowest possible $f_0$ and, hence, retains most physical
information at the low-frequency end.
\label{tab:f0}}
\begin{ruledtabular}
\begin{tabular}{lccc}
                                   & A1B3G3     & A1B3G5     & A3B3G3    \\
 Method                            & $f_0$ [Hz] & $f_0$ [Hz] & $f_0$ [Hz]  \\
\hline
 NP   ($R_{\rm ex}=1000 M_\odot$)  & $300$ & $300$           & $250$     \\
 RWZM ($R_{\rm ex}=1000 M_\odot$)  & $250$ & $400$           & $200$     \\
 CCE  ($R_\Gamma=1000 M_\odot$)    & $100$ & $100$           & $100$     \\
\end{tabular}
\end{ruledtabular}
\end{table}

We find that NP and RWZM extraction, which both operate at finite
radii, are subject to stronger drifts than the CCE method which
computes the waveforms gauge invariantly at future null infinity
$\scri^+$.  This is not surprising, given that the two former methods
both suffer from near-zone and gauge errors which typically lead to
low-frequency drifts in the waveform.  Hence, the strain $h^{\rm
CCE}_{e,+}$ as computed by CCE retains most of the physical
information at the low-frequency end of the spectrum with a cut-off at
$f_0 = 100\,\mathrm{Hz}$.  Unfortunately, even this value may not be
low enough, given that this frequency falls right into the band of
highest sensitivity of km-scale ground-based detectors
\cite{Abbott:2007kv, LIGO-sens-2010}.  Not being able to resolve the
low-frequency components is clearly a drawback of the curvature-based
extraction methods.

Fortunately, as we will discuss in more detail in
Sec.~\ref{sec:comparison-to-cce}, frequencies below $100\,\rm{Hz}$ do
not contribute significantly to the inferred theoretical signal-to-noise ratio (SNR) for
the models considered in this study or for the GW signal associated
with core bounce in slowly to moderately rapidly rotating core
collapse in general \cite{dimmelmeier:08}.  Hence, at least the CCE
method yields robust predictions for detection. A closer and more
detailed comparison between the waveforms computed with the various
methods will be performed in Sec.~\ref{sec:comparison-to-cce}.

In the following, we use the cut-offs $f_0$ as given in Table~\ref{tab:f0}
for the various models and extraction methods.

\subsection{Radial Dependence}
\label{sec:radial}

The physical gravitational strain $h$ scales $\propto 1/D$ with
distance $D$, provided it is computed in an asymptotic frame at large
distances from the source $D\rightarrow\infty$ (i.e.,~at astrophysical
distances).  At large asymptotic distances, we should therefore
observe $Dh=\rm{const}$.  Since NP extraction and the RWZM formalism
are evaluated at a finite distance from the source, they are both
subject to finite-radius errors and we will generally not exactly
observe $D h=\mathrm{const}$. Rather, we expect the signal to converge
with increasing extraction radius towards its asymptotic shape and
magnitude. In the context of vacuum binary black hole mergers, the
asymptotic behavior and finite-radius error of NP extraction has been
analyzed in Ref.~\cite{Pollney:2009ut}, where it is found that
extrapolations based on extractions at radii $R>300\,M$, where $M$ is
the total mass of the system, yield acceptable results.  In our case,
however, the finite-radius error contains the additional error arising
from non-zero matter content at the CCE world-tube locations and
NP/RWZM extraction spheres.

In Table~\ref{tab:rho}, we summarize the time-averaged matter
densities $\langle\rho\rangle$ at the various extraction spheres in
our models. For simplicity, we do not compute the extraction-surface-averaged 
matter density but simply report the equatorial density at
the various extraction spheres. The most compact model A3B3G3 has
$\langle\rho\rangle$ that are a factor of a few smaller at any given
radius than in the other models. Therefore, we expect the systematic
finite-radius error to be smallest in model A3B3G3.

In order to quantify the finite-radius error, we compute $D h_{+,e}$
at a succession of extraction spheres with increasing radii $R_{\rm
ex}= \{ 500\,M_\odot$, $700\,M_\odot$, $800\, M_\odot$,
$900\,M_\odot$, $1000\,M_\odot\}$ and evaluate the differences.  For
$R_{\rm ex}\geq 1000 M_\odot$, the spatial resolution of our
computational grid becomes too coarse for accurate wave extraction and
we limit our analysis to $R_{\rm ex}\leq 1000 M_\odot$
(Table~\ref{tab:rho}). Note that for a given model and extraction method, 
we use the same cut-off frequency 
$f_0$ for all extraction radii and world-tube locations.

In principle, we should extrapolate the waveforms as obtained at the
different extraction spheres to infinity. We observe, however, that
the differences at large radii are within our numerical errors (see
Sec.~\ref{sec:convergence}) and asymptote rapidly. Therefore, we
simplify the analysis by inspecting the behavior at successive radii
without extrapolating.  We will see in
Sec.~\ref{sec:comparison-to-cce} that this approach is justified. The
CCE method, which does not require any extrapolation, shows good
agreement with results obtained at finite radius within our numerical
errors.

\begin{table*}[t]
\caption{Time-averaged equatorial matter densities and their variations at extraction 
radii for the computed models. The first two columns 
report the extraction radii in code units $M_\odot$ and in cgs 
units, respectively.
All extraction surfaces are located on the fourth refinement level with a spatial resolution
of $\Delta x = 8 M_\odot$ ($\sim11.82\,\rm{km}$) and time resolution of $\Delta t = 1.6 M_\odot$ ($\sim7.9\times10^{-6}\,\rm{s}$).
\label{tab:rho}}
\begin{ruledtabular}
\begin{tabular}{ccccc}
 $R_{\rm ex}$ & $R_{\rm ex}$ & $\langle\rho\rangle$ A1B3G3 & $\langle\rho\rangle$ A1B3G5 & $\langle\rho\rangle$ A3B3G3 \\
 $[M_\odot]$  & [km]         & [g $\rm{cm}^{-3}$]          & [g $\rm{cm}^{-3}$]          & [g $\rm{cm}^{-3}$] \\
\hline
500  & 739  & $(1.2\pm0.3)\times10^{8}$ & $(1.6\pm0.4)\times10^{8}$ & $(8.6\pm0.2)\times10^{8}$ \\
700  & 1034 & $(2.3\pm0.4)\times10^{7}$ & $(2.6\pm0.5)\times10^{7}$ & $(1.3\pm0.3)\times10^{7}$ \\
800  & 1182 & $(8.6\pm1.2)\times10^{6}$ & $(9.3\pm1.2)\times10^{6}$ & $(3.5\pm0.9)\times10^{6}$ \\
900  & 1329 & $(3.0\pm0.4)\times10^{6}$ & $(3.3\pm0.4)\times10^{6}$ & $(5.5\pm1.6)\times10^{5}$ \\
1000 & 1477 & $(9.3\pm0.6)\times10^{5}$ & $(1.2\pm0.6)\times10^{5}$ & $(2.5\pm2.2)\times10^{4}$ \\
\end{tabular}
\end{ruledtabular}
\end{table*}

\begin{figure*}
  \includegraphics[width=1.\linewidth]{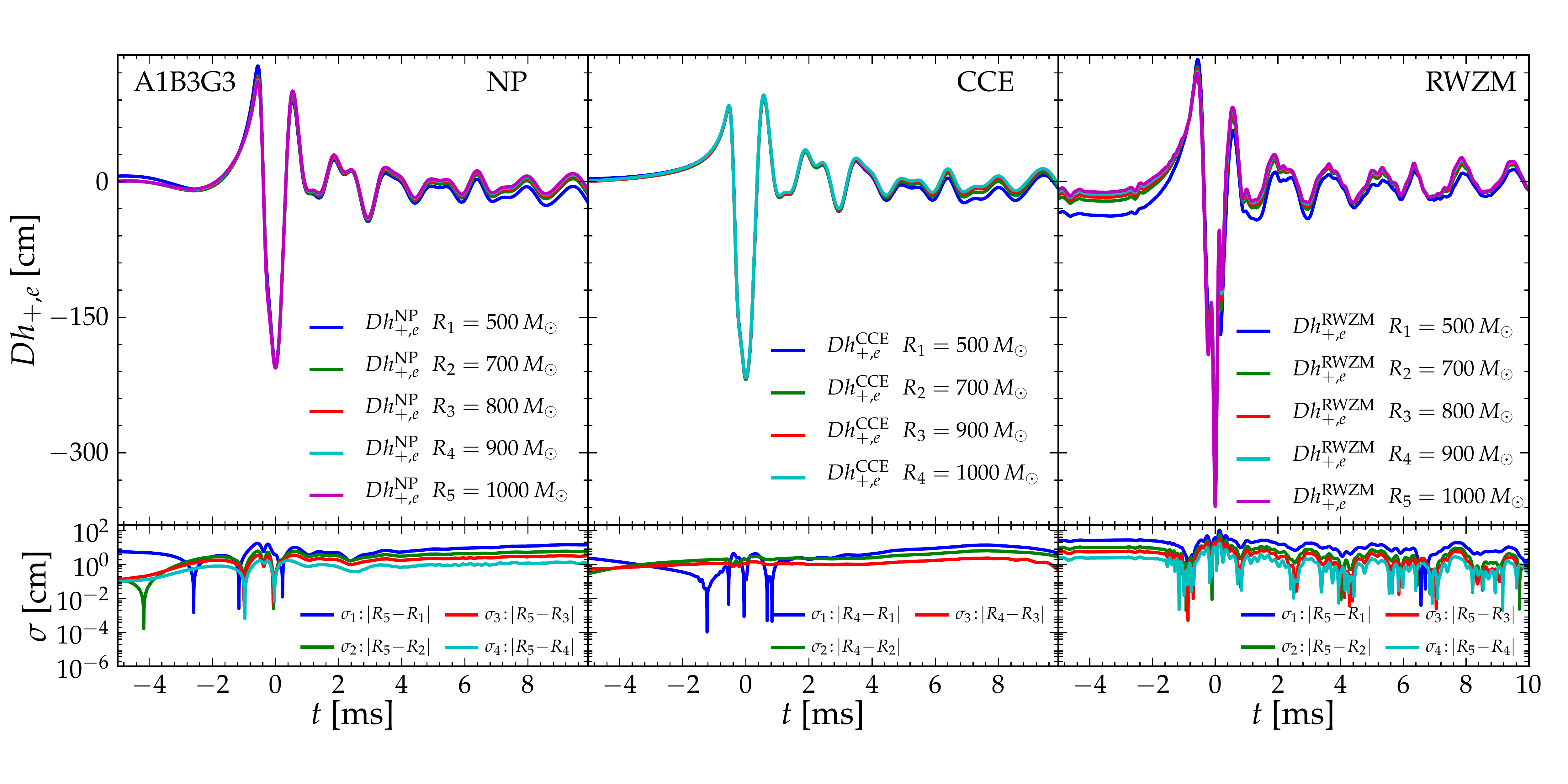}
  \vspace{-1.0cm}
  \caption{Top panel (from left to right): $Dh_{+,e}^{\rm NP}$ computed using NP extraction at radii $R_{\rm
      ex}=(500,700,800,900,1000)\,M_\odot$, $Dh_{+,e}^{\rm CCE}$ computed with CCE at $\scri^+$ using
    world-tube data at $R_{\Gamma}=(500,700,900,1000)\,M_\odot$, and $Dh_{+,e}^{\rm RWZM}$ computed using the RWZM formalism at radii
    $R_{\rm ex}=(500,700,800,900,1000)\,M_\odot$,
    all for model A1B3G3.  
    Bottom panels: Absolute difference $\sigma$ of the waves extracted at the various extraction radii and world-tube locations from those
    extracted at the outermost radius/location.
    The waveforms converge with increasing extraction radius and world-tube location.
    For NP extraction, we measure at the outermost
    detector sphere $R_{\rm ex}=1000\,M_\odot$ a
    maximum difference to the next closest detector $R_{\rm ex}=900\,M_\odot$
    of $\sigma_4=2\,\rm{cm}$, which corresponds to a percentage error
    of $\sim{1}\%$ relative to the maximum.
    For CCE, we measure a maximum difference of $\sigma_4=1.5\,\rm{cm}$, corresponding
    to a percentage error of $\sim{0.7}\%$ relative to the maximum.
    For RWZM extraction, we have $\sigma_4=14\,\rm{cm}$, corresponding to $\sim4\%$ relative to the maximum.
    We note that RWZM is subject to additional high-frequency features and also contains a spurious spike in the bounce peak.
    }
  \label{fig:radial_dep_A1B3G3}
\end{figure*}

\begin{figure*}
  \includegraphics[width=1.\linewidth]{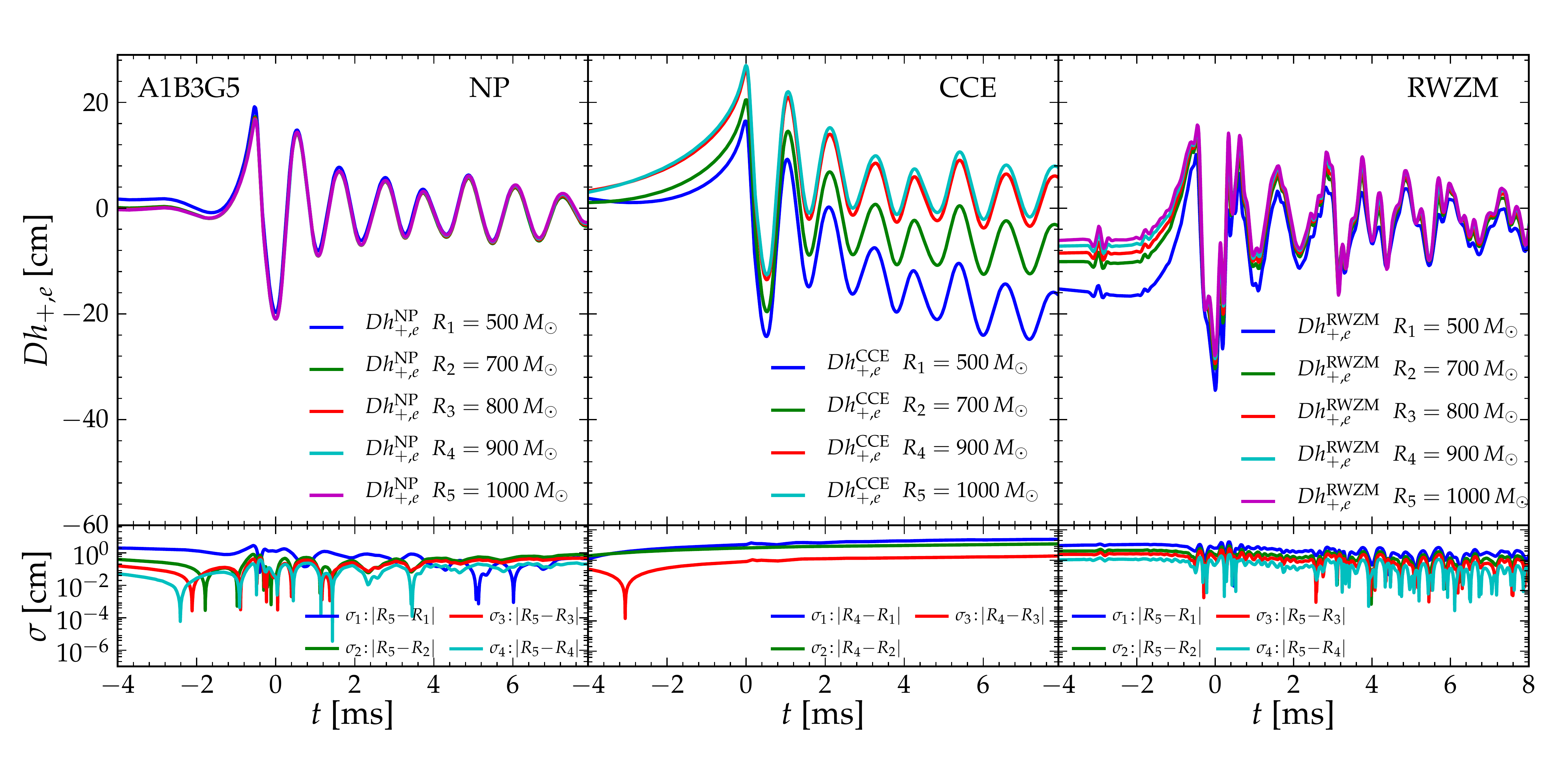}
  \vspace{-1.0cm}
  \caption{Radial dependence of waveforms computed for model A1B3G5. See caption of Fig.~\ref{fig:radial_dep_A1B3G3} for details.
    For NP extraction, we measure at the outermost
    detector sphere $R_{\rm ex}=1000\,M_\odot$ a
    maximum difference of $\sigma_4=0.4\,\rm{cm}$ to the next closest detector $R_{\rm ex}=900\,M_\odot$.
     This corresponds to a percentage error
    of $\sim{2}\%$ relative to the maximum.
    For CCE, we measure a maximum difference of $\sigma_4=1.6\,\rm{cm}$, corresponding
    to a percentage error of $\sim{6}\%$ relative to the maximum.
    For RWZM extraction, we have $\sigma_4=2.4\,\rm{cm}$, corresponding to $\sim8\%$ relative to the maximum.
    We note that RWZM is subject to strong additional high-frequency features. We also note that, while CCE shows for this model
    a larger error due to matter effects than finite-radius $\Psi_4$ extraction, it is more accurate at low frequencies (cf.~Table~\ref{tab:f0}).}
  \label{fig:radial_dep_A1B3G5}
\end{figure*}

\begin{figure*}
  \includegraphics[width=1.\linewidth]{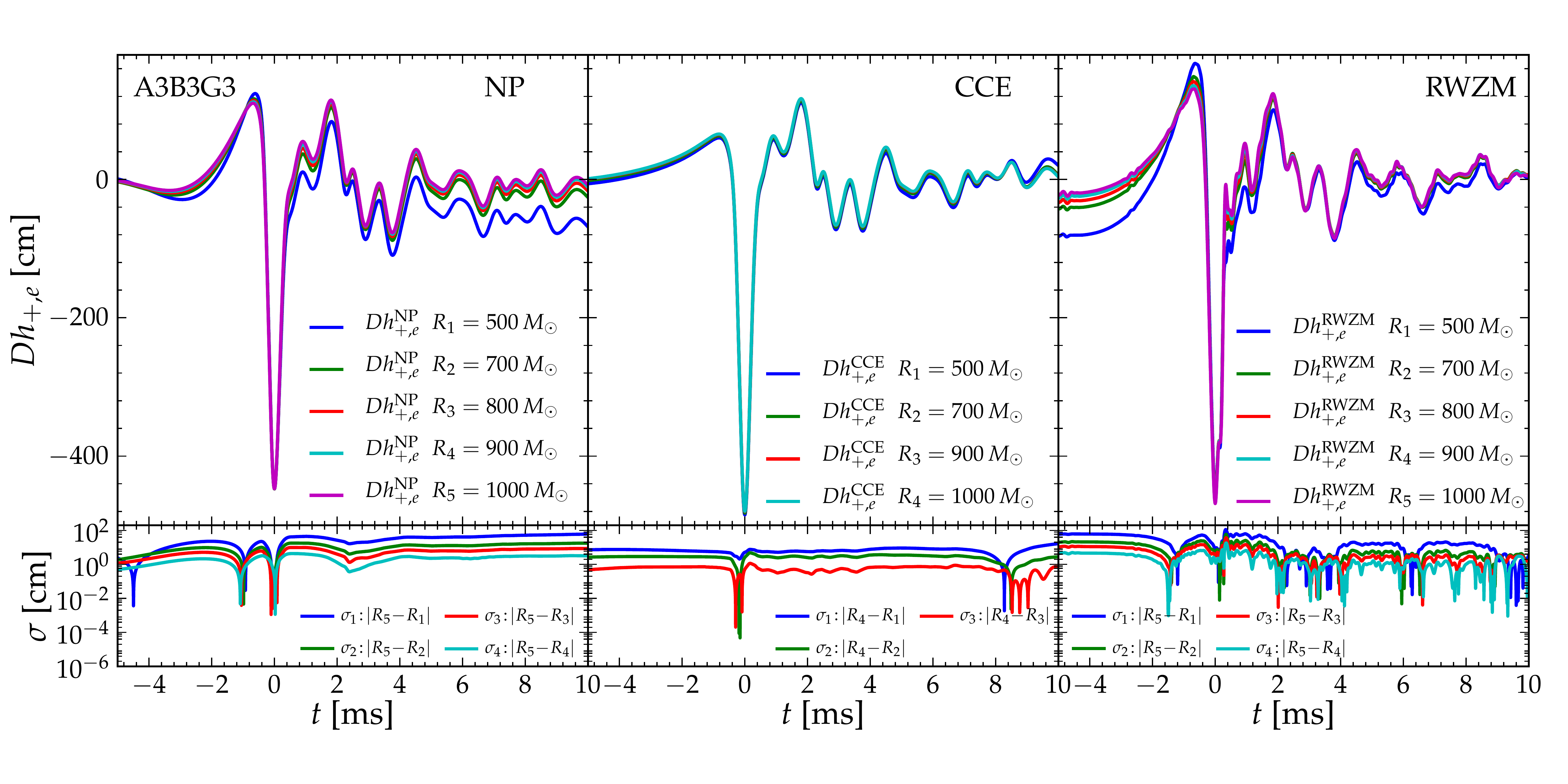}
  \vspace{-1.0cm}
  \caption{Radial dependence of waveforms computed for model A3B3G3. See caption of Fig.~\ref{fig:radial_dep_A1B3G3} for details.
    For NP extraction, we measure at the outermost
    detector sphere $R_{\rm ex}=1000\,M_\odot$ a
    maximum difference of $\sigma_4=2\,\rm{cm}$ to the next closest detector $R_{\rm ex}=900\,M_\odot$.
     This corresponds to a percentage error
    of $\sim{1}\%$ relative to the maximum.
    For CCE, we measure a maximum difference of $\sigma_4=1\,\rm{cm}$, corresponding
    to a percentage error of $\sim{0.2}\%$ relative to the maximum.
    For RWZM extraction, we have $\sigma_4=17\,\rm{cm}$, corresponding to $\sim4\%$ relative to the maximum.}
  \label{fig:radial_dep_A3B3G3}
\end{figure*}

\subsubsection{The NP Scalar $\Psi_4$}

In the upper left panels of
Figs.~\ref{fig:radial_dep_A1B3G3},~\ref{fig:radial_dep_A1B3G5},
and~\ref{fig:radial_dep_A3B3G3}, we show $D h_{+,e}^{\rm NP}$ as
computed from the NP scalar $\Psi_4$ for model A1B3G3, A1B3G5 and
A3B3G3, respectively. In the bottom panel, we show the absolute
differences $\sigma$ between $D h_{+,e}^{\rm NP}$ at $R_{\rm ex}$ from
the reference distance at $R_{\rm ex}=1000\,M_\odot$. In an ideal
asymptotic frame, all curves would line up exactly.  This is not the
case in practice. We notice that the curves asymptote with increasing
extraction radius, i.e.,~the differences $\sigma$ between two
successive extraction spheres converge to zero. This behavior shows
that our extraction radii, albeit rather close to (and even inside)
the star, lead to finite-radius errors for the waveforms computed from
the NP scalar $\Psi_4$ that are below the discretization errors
(cf.~Sec.~\ref{sec:convergence}).  We measure a maximum absolute
difference in amplitude of $\sigma \lesssim 2\,\rm{cm}$ between the
two outermost extraction spheres of any model (see lower panels of
Figs.~\ref{fig:radial_dep_A1B3G3},~\ref{fig:radial_dep_A1B3G5},
and~\ref{fig:radial_dep_A3B3G3}). This corresponds to a relative error
of no more than $\sim 2\%$ when compared to the peak amplitudes.

Note that we had to cut-off low frequencies (Table~\ref{tab:f0}) in
order to remove spurious non-linear drifts. This restricts our
analysis to frequency components above the cut-off frequency $f_0$.
Experiments show that for larger extraction radii, artificial
low-frequency components become less prominent. Hence, extracting at
greater radius would allow us to decrease $f_0$, but is presently too
computationally demanding to be possible for production simulations.

\subsubsection{The Regge-Wheeler-Zerilli-Moncrief Formalism}

The RWZM variables are computed as perturbations on an assumed fixed
background spacetime.  
In an asymptotic frame at large distances from the source, they
are independent of radius.  As in the case for NP extraction, this
is not achieved in practice, but the residual errors should 
converge with increasing extraction radius.
We measure a relative difference in amplitude between the two
outermost detector spheres of $\sigma \lesssim 17\, \rm{cm}$ for all models
(see the right panels of
Figs.~\ref{fig:radial_dep_A1B3G3},~\ref{fig:radial_dep_A1B3G5}
and~\ref{fig:radial_dep_A3B3G3}). This corresponds to a
relative error of $\lesssim8\%$ when compared to the maximum
amplitudes and is significantly larger than what we find for NP extraction.

In addition, the RWZM method produces high-frequency variations in the
waveform at core bounce and similar high-frequency features in the
ringdown phase that are not seen in GWs extracted with the other
methods. These features do not appear to converge with increasing
radius; at least not at radii accessible to our simulations. They are
particularly manifest in GWs of models producing weak signals, e.g.,
in the signal emitted by model A1B3G5 of our model set
(Fig.~\ref{fig:radial_dep_A1B3G5}). Furthermore, and most pronounced
in model A1B3G3's waveform, a large spike during core bounce is
visible in the RWZM result, but is not produced by any of the other
methods (see Fig.~\ref{fig:radial_dep_A1B3G3} and the comparison in
the upper panel of Fig.~\ref{fig:comp_h_A1B3G3}).

In order to investigate the cause of the differences seen with RWZM
extraction, we perform a range of test calculations. These include
(\emph{i}) using two additional \emph{independent} implementations of
the RWZM method, one assuming a Minkowski background, the other using
a generalization of the RWZM approach \cite{Pazos:2006kz}, (\emph{ii})
performing a computationally very expensive simulation with extended
grids, allowing for RWZM extraction at $R_\mathrm{ex} =
3000\,M_\odot$, (\emph{iii}) performing simulations with up to a
factor of $2$ higher resolution and modified mesh refinement boundary
locations, and (\emph{iv}) changing the spacetime gauge conditions,
including exponential damping of the evolution of the coordinate shift
at large radii near the extraction spheres.

None of the above tests leads to any significant change of the RWZM
result. This brings us to the conclusion that the high-frequency
features observed in RWZM waveforms are systematic problems tied, most
likely, to the particular perturbative nature of the RWZM scheme.  One
notable difference of the RWZM formalism from the other methods is the
procedure of projecting out the spherical background geometry (e.g.~Eq.~\ref{eq:H2lm}).  This
can result in very small values for the aspherical perturbation
coefficients that are prone to numerical noise and cancellation
effects.  The RWZM approach may therefore be less suitable for the
extraction of the generally weak GW signals emitted in core collapse.

\subsubsection{Radial Dependence on World-Tube Location for CCE}

The CCE method uses metric data from a time succession of
finite-radius coordinate spheres, the world-tube, as inner boundary
data for the evolution of the gravitational field out to $\scri^+$.
In vacuum, the method does not depend on the particular choice of any
given world-tube radius \cite{Reisswig:2009us, Reisswig:2009rx}.
However, the presence of matter at the world-tube location leads to a
systematic error and imposes an artificial dependence of the waveforms
computed at $\scri^+$ on the world-tube location.  We plot in the
center two panels of Figs.~\ref{fig:radial_dep_A1B3G3},
\ref{fig:radial_dep_A1B3G5}, and \ref{fig:radial_dep_A3B3G3} the
waveforms obtained at $\scri^+$ using different world-tube radii as
inner boundaries for the characteristic evolution. The center bottom
panel of these figures depicts the absolute difference $\sigma$
between the waveforms from the outermost world-tube radius at
$R_{\Gamma}=1000\,M_\odot$ and the waves from each of the smaller
world-tube radii.  It is apparent that the differences between the
outermost two world-tube radii is always smallest, with absolute
differences $\sigma < 1.6\, \rm{cm}$ for all models and all times, and
with an error relative to the maximum amplitude of $\lesssim1\%$ for
models A1B3G3 and A3B3G3, and $\sim6\%$ for model A1B3G5. For the
latter model, we also notice strong drifts at the innermost world-tube
radii. Systematic errors due to the presence of matter can therefore
become significant when the signal is weak and the density large. Note
that the strong drift at the innermost world-tubes may be removed by
an increased FFI cut-off frequency of $f_0=150\,\rm{Hz}$ in this case.
Since we would like to retain as much physical information as
possible, and since the outermost world-tube location permits
$f_0=100\,\rm{Hz}$, we have chosen this value for all world-tubes.
Generally, the lower FFI cut-off frequency $f_0$ induces a more
sensitive radial behavior with respect to low-frequency drifts.  Since
CCE permits a lower $f_0$ than NP extraction, the radial variations
are slightly larger for CCE in model A1B3G5 when compared to the
radial variations of NP extraction (cf.~bottom panels of
Fig.~\ref{fig:radial_dep_A1B3G5}).  In the other models, we
find smaller radial variations between the two outermost CCE
world-tubes than between the two outermost NP extraction spheres
(cf.~bottom panels of Figs.~\ref{fig:radial_dep_A1B3G3} and
\ref{fig:radial_dep_A3B3G3}), even though $f_0$ is smaller for CCE
than for NP extraction in these models as well.  This indicates that
the remaining systematic non-zero matter error in CCE is not as
important as the additional near-zone errors and gauge ambiguities
inherent to NP extraction.

\subsection{Comparison}

\subsubsection{Comparison with Cauchy-Characteristic Extraction}
\label{sec:comparison-to-cce}

\begin{figure}
  \includegraphics[width=1.\linewidth]{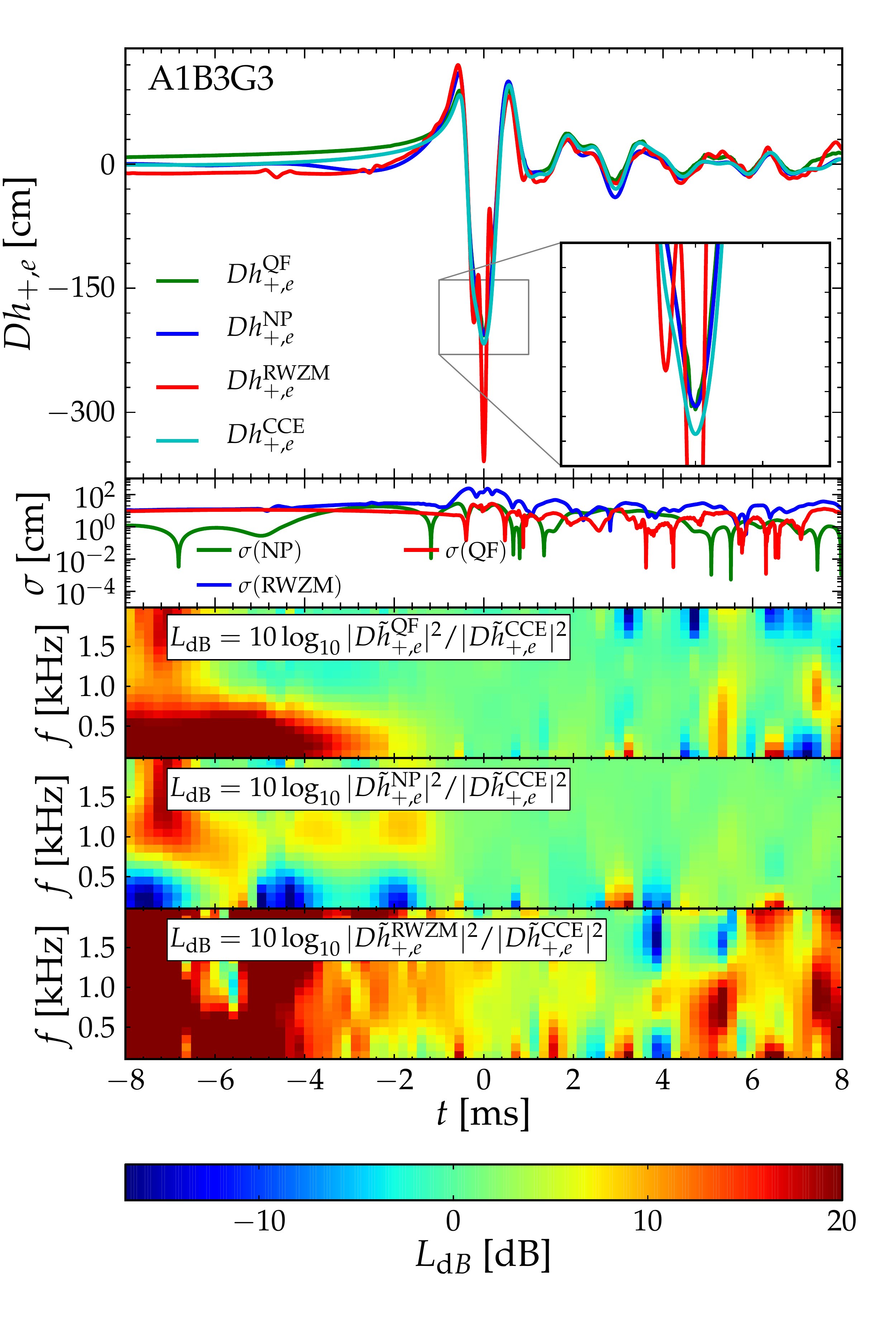}
  \vspace{-1.0cm}
   \caption{Comparison of waveform amplitudes $Dh_{+,e}$, their
           absolute differences $\sigma$ from CCE waveforms, and
           spectrograms of the power ratio $L_{\rm dB}$ between
           waveforms obtained from each extraction method and
           waveforms obtained via CCE for model A1B3G3. If $L_{\rm
           dB}=0$, the waveform of the particular extraction method
           yields equal power per time and frequency bin with respect to that
           obtained with CCE. This is indicated by green colors. Blue
           colors indicate less power, and red colors indicate more
           power. See text for details.  NP extraction and the
           quadrupole approximation yield remarkable agreement with
           CCE at frequencies below $2\,\mathrm{kHz}$ at and after
           core bounce, while the RWZM formalism yields a spurious
           spike during core bounce and generally contains artificial
           high-frequency oscillations. This also becomes clear in the
           spectrograms of the power ratio between RWZM formalism and
           CCE since $L_{\rm dB} > 0$ over a wide range of time and
           frequencies (\textit{bottom panel}).  Prior to core bounce
           $-8\,\rm{ms} < t < -1\,\rm{ms}$, NP extraction results in
           less power compared to CCE, while the QF yields more
           power. }
  \label{fig:comp_h_A1B3G3}
\end{figure}

\begin{figure}
  \includegraphics[width=1.\linewidth]{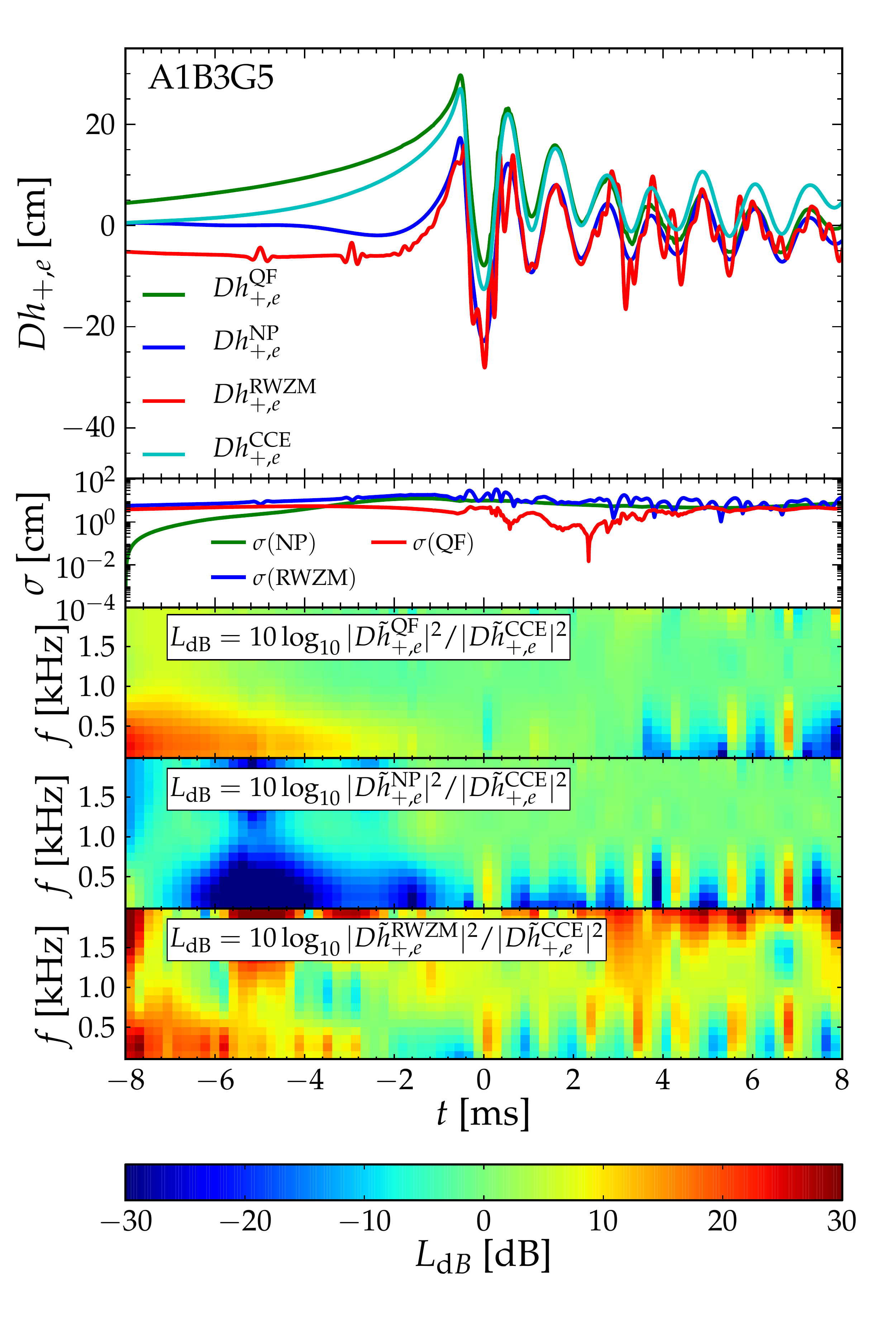}
  \vspace{-1.0cm}
  \caption{Comparison of waveform amplitudes $Dh_{+,e}$, their
           absolute differences $\sigma$ from CCE waveforms, and
           spectrograms of the power ratio $L_{\rm dB}$ between
           waveforms obtained from each extraction method and
           waveforms obtained via CCE for model A1B3G5. If $L_{\rm
           dB}=0$, the waveform of the particular extraction method
           yields equal power per time and frequency bin with respect to that
           obtained with CCE. This is indicated by green colors. Blue
           colors indicate less power, and red colors indicate more
           power. See text for details.  The waveforms from NP
           extraction and quadrupole approximation agree well at
           frequencies below $2\,\mathrm{kHz}$ at and after core
           bounce, while the RWZM formalism is subject to artificial
           high-frequency oscillations. The spectrograms of the power
           ratio between waveforms from the RWZM formalism and CCE in
           the bottom panel further support this, since $L_{\rm dB}>0$
           in this case.  Prior to core bounce $-8\,\rm{ms} < t <
           -1\,\rm{ms}$, NP extraction results in less power compared
           to CCE, while the QF yields more power.  }
  \label{fig:comp_h_A1B3G5}
\end{figure}

\begin{figure}
  \includegraphics[width=1.\linewidth]{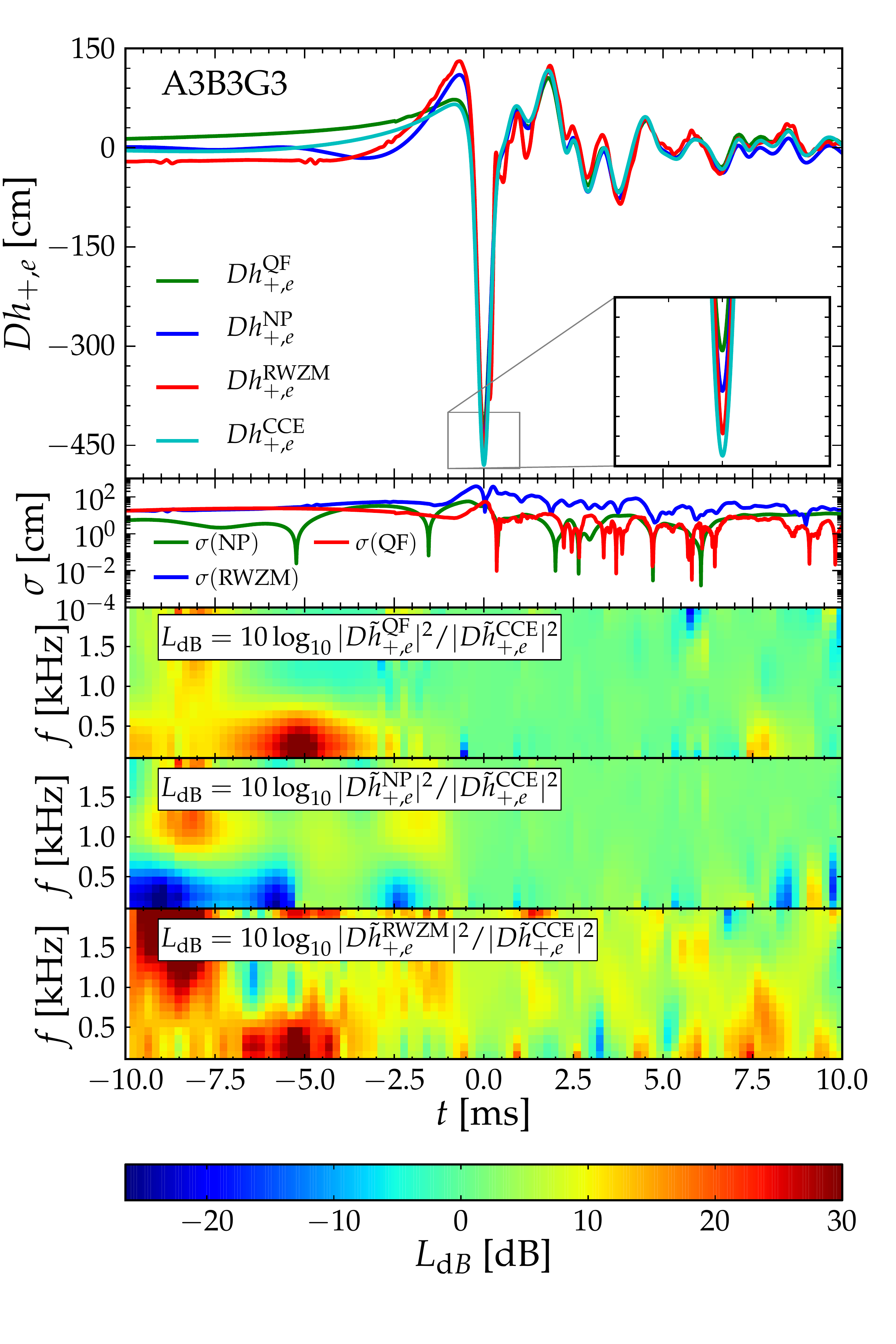}
  \vspace{-1.0cm}
  \caption{Comparison of waveform amplitudes $Dh_{+,e}$, their
           absolute differences $\sigma$ from CCE waveforms, and
           spectrograms of the power ratio $L_{\rm dB}$ between
           waveforms obtained from each extraction method and
           waveforms obtained via CCE for model A3B3G3. If $L_{\rm
           dB}=0$, the waveform of the particular extraction method
           yields equal power per time and frequency bin with respect to that
           obtained with CCE. This is indicated by green colors. Blue
           colors indicate less power, and red colors indicate more
           power. See text for further explanation.  The waveforms
           from NP extraction and from the quadrupole approximation
           agree well below $2\,\mathrm{kHz}$, while the waveform from
           the RWZM formalism contains artificial high-frequency
           oscillations thus leading to a higher power emission than
           waveforms obtained via CCE (\textit{bottom panel}).  Prior
           to core bounce $-10\,\rm{ms} < t < -2\,\rm{ms}$, NP
           extraction results in less power compared to CCE, while the
           QF yields more power.  }
  \label{fig:comp_h_A3B3G3}
\end{figure}

CCE yields waveforms that contain the least amount of systematic errors
compared to the other extraction methods considered in this work.  It is
completely gauge invariant and is free of near-zone errors.  As found
in Sec.~\ref{sec:int}, it is the only curvature-based method that
captures most of the low-frequency band.  Furthermore, as found in
Sec.~\ref{sec:radial}, the remaining error due to the non-zero
stress-energy tensor is small.  We therefore use the waveforms
obtained with CCE as a benchmark.  In Figs.~\ref{fig:comp_h_A1B3G3},
\ref{fig:comp_h_A1B3G5} and \ref{fig:comp_h_A3B3G3}, we examine the
waveforms as obtained by the various extraction methods, i.e., NP,
RWZM, QF, and CCE for each model. In each figure, the top panel
displays the amplitudes of the waves, $D h_{+,e}$, with the panel
right below showing the absolute differences $\sigma$ of each
extraction method from CCE\@.  It is apparent that in all cases the RWZM
formalism yields the largest differences from CCE\@.  As discussed in
Sec.~\ref{sec:radial}, the RWZM formalism is subject to high-frequency
noise and yields a large spurious spike at core bounce, most
pronounced in model A1B3G3.

The quadrupole approximation and NP extraction only lead to small
differences to CCE, in particular at core bounce, but also in the
ring-down part of the waveform.  Moreover, it is surprising that the
quadrupole formalism yields agreement so remarkably close to the
results obtained via CCE, given the rather simplistic nature of the
QF\@.  Quantitatively, when compared to the waveforms obtained via
CCE, we find for model A1B3G3 that the waves obtained via the
quadrupole approximation lead to smaller core-bounce peaks with
differences up to $\sim10\, \rm{cm}$ ($\sim5\%$).  For the same model,
NP extraction results in core-bounce peaks that are smaller as well,
with differences of $\sim5\%$.  Note, however, that in the NP
waveform, the first positive peak prior to bounce is much larger
$\sim31\%$ than what is predicted by CCE.  The QF result, on the
other hand, agrees much better with CCE at this peak ($\sim 5\%$
overprediction).

For model A3B3G3, we find that at the bounce peak the QF (NP)
amplitudes are $\sim 11\%$ ($\sim7\%$) smaller than the CCE
prediction.  For this model, the first positive peak before bounce is
overpredicted by NP by $\sim69\%$, while the QF yields an
overprediction of only $\sim11\%$ compared to CCE.

A separate treatment is necessary for model A1B3G5.  As briefly
discussed in Sec.~\ref{sec:int}, the physical low-frequency components
are filtered out by FFI in curvature-based extraction methods\@.  The
waveform of model A1B3G5 is most affected by this: Any physical
low-frequency modulations or offsets are removed. Hence, the waveforms
from curvature-based extraction are shifted downwards with respect to
the QF waveform that does not require filtering.  This shift leads to
large absolute differences in the peak amplitudes by as much as
$\sim-10\,\rm{cm}$ at the first and second peaks when compared to the
QF waveform, yielding relative differences by as much as $\sim90\%$ at
core bounce.  Constant (or nearly constant) offsets in the waveforms,
however, are not visible to GW detectors.  In order to get a better
measure of the differences between the various extraction methods, we
therefore compare the \emph{change} in amplitude $\delta
Dh_{+,e}=|Dh_{+,e}^{t_1}-Dh_{+,e}^{t_2}|$ between first and second
peak occurring at time $t_1$ and $t_2$, respectively.  Compared to
CCE, the change $\delta Dh_{+,e}^{\rm QF}$ measured in the quadrupole
waves is $\sim7\%$ smaller and $\delta Dh_{+,e}^{\rm NP}$ of the NP
waveform is $\sim1\%$ larger.

Since a hypothetical matched-filtering GW search would be sensitive 
to differences in phase, an important
aspect is the phase relation of the waveforms.  A
measure for the phase is given by the time lag between successive wave
peaks, and we label the time of the peaks by $t_n=t_1,\, t_2,\, t_3,
\ldots$ in their temporal order of occurrence.  Note that in all
cases, we have aligned the waveforms at the second peak occurring at
$t_2\equiv0\,\rm{ms}$ so that the time lag at this peak is zero for
all methods and all models.  We therefore measure the time lag $\delta
t_{n,2}$ for a peak $t_{n\neq2}$ relative to the second peak, e.g., we
measure $\delta t_{1,2}=|t_2-t_1|$.  By comparing a number of time
lags $\delta t_{n,2}$, we generally find that NP extraction, CCE, and
QF produce the same phasing with an error of less than
$\sim\pm0.05\,\rm{ms}$ for all peaks and all models.  The differences
in the time lags between successive wave peaks computed from the
various extraction methods are therefore close to our time resolution
of $\Delta t \approx 8\times10^{-3}\,\rm{ms}$.  Note that the RWZM
formalism is excluded from this analysis, since the additional
high-frequency components make it hard to clearly identify the times
of the maxima.  Visual inspection suggests a phase error for
RWZM comparable to the other methods, provided high-frequency
contributions are ignored.

In the bottom three panels of Figs.~\ref{fig:comp_h_A1B3G3},
\ref{fig:comp_h_A1B3G5} and \ref{fig:comp_h_A3B3G3}, we plot
spectrograms\footnote{The spectrograms are made up of 100 time bins of
$0.2\,\rm{ms}$ each and use a Hann window with a width of $2\,
\rm{ms}$.}  of the power ratios $L_{\rm dB}=10\log_{10}(P_1 / P_0)$.
Here, the power spectrum $P_1=|\tilde{h}(f)|^2$ is computed from the
quadrupole approximation, NP extraction, and the RWZM formalism,
respectively.  The power spectrum $P_0=|\tilde{h}_{\rm CCE}(f)|^2$ is
computed from the waves obtained via CCE\@.  At a given time $t$, the
power spectra $P_{0,1}$ are obtained from the short-time Fourier
transform of the strain over a time window of $2\,\rm{ms}$ centered at
$t$.  Thus, $L_{\rm dB}$ measures the power ratio per time and
frequency bin of all extraction methods relative to the CCE method.
If $L_{\rm dB}=0$, the extraction method emits equal power per time
and frequency bin and, hence, is equivalent to the waves from CCE\@.
This is indicated by green colors. Red regions indicate that waves
obtained with the corresponding extraction method emit more power
than the CCE waves during that time and frequency bin; blue indicates
less power.

By inspection of the spectrograms, it is apparent that at core bounce,
the NP method, and also the quadrupole approximation predict waves
carrying roughly equal powers with respect to the waveforms obtained
via CCE in a time interval of $[-4\,\rm{ms},4\,\rm{ms}]$ and over the
entire frequency range.  Furthermore, we observe that in the prebounce
phase $[-10\,\rm{ms},4\,\rm{ms}]$, the quadrupole waves emits more
power in low frequencies than the CCE waves, and the NP extracted
waveforms emit less power.  This is largely an effect of the different
cut-off frequencies introduced in the integration of the waves
obtained from the curvature-based methods (Table~\ref{tab:f0}).

We also find from the spectrograms that the RWZM extraction always
deviates strongest from all other methods, primarily because its GWs
contain spurious additional high-frequency components
(cf.~bottom panels of Figs.~\ref{fig:comp_h_A1B3G3},
\ref{fig:comp_h_A1B3G5} and \ref{fig:comp_h_A3B3G3}). This is clearly
visible in the spectrograms, which show additional red ``speckles'' of
higher emitted power throughout the wavetrain and frequency band.

As opposed to the quadrupole approximation, all curvature-based
extraction methods are filtered below a frequency $f_0$
(Sec.~\ref{sec:int}).  It is crucial to know whether the missing low
frequencies can spoil the detectability of the GWs.
To gauge the influence of low-frequencies on the theoretical signal
strength in GW detectors, we compute the theoretical optimal SNR for
the quadrupole waveform once including low-frequency contributions and
once artificially cutting them off.  The SNR is given by
\cite{Flanagan:1997sx}
\begin{equation} \label{eq:snr}
\rho_S^2 = 4\, \mathrm{Re} \int_0^\infty df \frac{|\tilde{h}(f)|^2}{S_h(f)}\,,
\end{equation}
where $\tilde{h}$ is the Fourier transform of the strain $h$ as
measured at a distance $D=10\,\rm{kpc}$ to the source, and $S_h$ is
the one-sided noise power spectral density for a given detector, i.e.,
the detector sensitivity function.

Table~\ref{tab:snr} lists for all models the theoretical optimal SNRs
of waves extracted with the quadrupole formalism for the LIGO
\cite{Abbott:2007kv,LIGO:2010cg} detector and for the zero-detuning high-power
configuration of advanced LIGO \cite{LIGO-sens-2010}.  By
cutting off at $f_0=100\, \rm{Hz}$ (``QF, $f_0=100$ Hz''), we observe
no significant reduction in SNR for models A1B3G3 and A3B3G3
and, hence, no loss of detectable GW information in rotating core
collapse for these models.  
Model A1B3G5's type-III waveform, on the other hand, has
low-frequency components of high relative strength.  In this model, a
cut-off at $f_0=100\,\rm{Hz}$ already leads to a SNR reduction by
$\sim17\%$ ($\sim23\%$) in LIGO (advanced LIGO).

When using $100\,\rm{Hz}<f_0< 300$
Hz as a cut-off, we find a reduction in SNR as large as $\sim10-30\%$ for all
models.  Comparing to the cut-off frequencies required for FFI
(Table~\ref{tab:f0}), this indicates that the finite-radius
curvature-based NP and RWZM methods suffer from their inability to
properly resolve frequencies in the range
$100\,\rm{Hz}<f<300\,\rm{Hz}$.  

As shown in Table~\ref{tab:snr}, the theoretical optimal SNRs as
computed from the waves obtained for models A1B3G3 and A3B3G3 via CCE
are essentially unaffected by low-frequency removal.  Instead, they
yield a slightly increased SNR compared to the QF result. This is due
mostly to the higher amplitudes of the CCE waves at core bounce.
Since low-frequency components are missing, model A1B3G5 is subject to
a loss in SNR.  CCE yields SNRs for LIGO and advanced LIGO that are
identical to those obtained with QF with a cut-off at
$f_0=100\,\rm{Hz}$ (``QF, $f_0=100\,\rm{Hz}$'').

The RWZM method always overpredicts the SNR by $\sim40$-$100\%$.  The
reason is apparent from the spectrograms in the bottom panels of
Figs.~\ref{fig:comp_h_A1B3G3}, \ref{fig:comp_h_A1B3G5}, and
\ref{fig:comp_h_A3B3G3}, in which spurious additional high-frequency
components lead to artificially high GW power and  a
corresponding overestimate of the SNR.

The waves computed with NP extraction show a high degree of agreement
in SNR with the waves obtained with the QF.  This, however, is
misleading, since there are two balancing effects: (i) NP
extraction predicts a higher amplitude in the first peak prior to core
bounce, which would yield a larger SNR, but (ii), in NP extraction we must
cut off frequencies $f<300\,\rm{Hz}$, which artificially reduces the
SNR\@.

Finally, we address the question of whether the waves obtained with the
various extraction methods are within the tolerance for detection in a
(hypothetical) matched-filtering GW data analysis of LIGO and advanced
LIGO\@.  Ideally, waveforms for the same model lead to a detection of
the same model parameters and hence should not be distinguishable
within a given threshold.

\begin{table}[t]
\caption{SNR $\rho_S$ according to Eq.~(\ref{eq:snr}) for all models
and all extraction methods at a distance $D=10\,\rm{kpc}$ for LIGO
(top) and advanced LIGO (bottom).  Note that we take into account only
those frequencies above a given $f_0$ that have been determined to be
reliable (see Table~\ref{tab:f0}).  Since the quadrupole formalism is
robust at low frequencies, we can gauge the influence of the neglected
frequencies by additionally computing the SNR for the quadrupole
waveform with the same low-frequency cut-off. For LIGO, we find no
loss in SNR. \change{[]} For advanced LIGO the loss is at $\sim1\%$. An
exception is model A1B3G5 where the low-frequency components 
contribute significantly to the total emission.
\label{tab:snr}}
\begin{ruledtabular}
\begin{tabular}{lrrrrr}
  Method                        & A1B3G3 & A1B3G5 & A3B3G3 \\
                                & $\rho_S$ & $\rho_S$ & $\rho_S$ \\
\hline
  \underline{\textit{LIGO}}                          & & & \\
  QF                            & $3.5$  & $0.6$ & $7.8$ \\
  QF ($f_0=100$ Hz)             & $3.5$  & $0.5$ & $7.8$ \\
  NP ($R_{\rm ex}=1000M_\odot$) & $3.1$  & $0.4$  & $7.4$ \\
  RWZM ($R_{\rm ex}=1000M_\odot$) & $6.9$ & $0.7$ & $17.0$ \\
  CCE ($R_\Gamma=1000M_\odot$)    & $3.8$  & $0.5$ & $8.5$ \\
\hline
  \underline{\textit{advanced LIGO}}               & & & \\
  QF                          & $49$  & $9$ & $95$ \\
  QF ($f_0=100$ Hz)           & $49$  & $7$ & $94$ \\
  NP ($R_{\rm ex}=1000M_\odot$) & $49$  & $6$ & $96$ \\
  RWZM ($R_{\rm ex}=1000M_\odot$) & $105$ & $12$ & $209$ \\
  CCE ($R_\Gamma=1000M_\odot$     & $52$ & $7$ & $103$ \\
\end{tabular}
\end{ruledtabular}
\end{table}

\begin{table}[t]
 \caption{Mismatch $\mathcal{M}_{\rm mis}$ according to
   Eq.~(\ref{eq:mismatch}) for all models between CCE and the other
   extraction methods for the LIGO (top) and advanced LIGO detector
   (bottom).  Note that we take into account only those frequencies
   above a given $f_0$ that have been determined to be reliable (see
   Table~\ref{tab:f0}). The waveforms from the quadrupole
   approximation yield the smallest mismatch to the waves from CCE,
   except in model A1B3G3. But note that the quadrupole approximation
   yields waveforms that allow the inclusion of lower frequencies than
   all other methods and hence allow the computation of the mismatch
   over a greater frequency range (only limited by the cut-off
   frequency of CCE) which introduces a small bias.
\label{tab:mismatch}}
\begin{ruledtabular}
\begin{tabular}{lrrrrr}
  Method                        & A1B3G3 & A1B3G5 & A3B3G3 \\
                                & $\mathcal{M}_{\rm mis}$ & $\mathcal{M}_{\rm mis}$ & $\mathcal{M}_{\rm mis}$ \\ 
\hline
  \underline{\textit{LIGO}}                          & & & \\
  NP ($R_{\rm ex}=1000M_\odot$) & $5\times10^{-3}$  & $19\times10^{-3}$ & $8\times10^{-3}$ \\
  RWZM ($R_{\rm ex}=1000M_\odot$) & $12\times10^{-3}$ & $38\times10^{-3}$ & $5\times10^{-3}$ \\
  QF                                  & $6\times10^{-3}$  & $13\times10^{-3}$ & $2\times10^{-3}$ \\
\hline
  \underline{\textit{advanced LIGO}}               & & & \\
  NP ($R_{\rm ex}=1000M_\odot$) & $3\times10^{-3}$  & $12\times10^{-3}$ & $7\times10^{-3}$ \\
  RWZM ($R_{\rm ex}=1000M_\odot$) & $14\times10^{-3}$ & $48\times10^{-3}$ & $7\times10^{-3}$ \\
  QF                                  & $4\times10^{-3}$  & $7\times10^{-3}$ & $2\times10^{-3}$ \\
\end{tabular}
\end{ruledtabular}
\end{table}

The (dis)agreement of waveforms obtained from different methods can be
quantified by the mismatch (see,
e.g.~\cite{Owen:1995tm,Damour:1997ub})
\begin{equation} \label{eq:mismatch}
\mathcal{M}_{\rm mis} = 1-\mathcal{M}\,,
\end{equation}
where the \textit{best} match $\mathcal{M}$ is given by
\begin{equation}
\mathcal{M} = \max_{t_0}\max_{\phi_1}\max_{\phi_2} \mathcal{O}[h_1,h_2]\,,
\end{equation}
which involves a maximization over time of arrival $t_0$ and the two
phases $\phi_1$ and $\phi_2$ of the two wave signals $h_1$ and $h_2$,
respectively.  The overlap $\mathcal{O}$ between two waveforms is
given by
\begin{equation}
\mathcal{O}[h_1,h_2]:=\frac{\langle h_1 | h_2 \rangle}{\sqrt{\langle h_1 | h_1 \rangle\langle h_2 | h_2 \rangle}}\,,
\end{equation}
with the detector-noise weighted scalar product
\begin{equation}
\langle h_1 | h_2 \rangle = 4\, \mathrm{Re} \int_0^\infty df \frac{\tilde{h}_1(f)\tilde{h}_2^*(f)}{S_h(f)}\,.
\end{equation}
A mismatch $\mathcal{M}_{\rm
mis}$ of zero indicates that waveforms $h_1$ and $h_2$ are identical.
Conversely, a mismatch of $\mathcal{M}_{\rm mis}=1$ indicates that the
waveforms are completely different.  In Table~\ref{tab:mismatch}, we
list the mismatches between each of the extraction methods and the CCE
method for all models. Note that we compute the mismatch starting from
$f_0 = \max\{f_0^{(1)}, f_0^{(2)}\}$, where $f_0^{(1)}$ and
$f_0^{(2)}$ are the lower cut-off frequencies as listed in
Table~\ref{tab:f0} for the waveforms $h_1$ and $h_2$, respectively,
since we do not trust waveforms below their value of $f_0$.  We find
that in all cases, the quadrupole approximation agrees best with
waveforms obtained via CCE with mismatches to within $1\%$ or better:
We find a mismatch in the quadrupole waveforms to CCE of $0.6\%$
($0.4\%$) for LIGO (advanced LIGO) for model A1B3G3, $1.3\%$ ($0.7\%$)
for LIGO (advanced LIGO) for model A1B3G5, and $0.2\%$ ($0.2\%$) for
LIGO (advanced LIGO) for model A3B3G3.  As reported in
Table~\ref{tab:mismatch}, NP extraction leads to slightly larger
mismatches, due mainly (i) to less emitted power in the low-frequency
band at and above this method's cut-off frequency, and (ii) to higher
emitted power in the first wave peak (cf.~spectrograms in second-lower
panels of Figs.~\ref{fig:comp_h_A1B3G3}, \ref{fig:comp_h_A1B3G5}, and
\ref{fig:comp_h_A3B3G3}).  The RWZM formalism performs worst for
models producing weak signals, e.g., model A1B3G3 and, in particular,
model A1B3G5. This is due primarily to the artificial high-frequency
components produced by this method.  Note that the mismatch discussed
above depends on the cut-off frequency $f_0$.  As a result, the range
of frequencies contributing to the mismatch calculation is greatest
for CCE-QF and smallest for CCE-RWZM\@. Hence, a full unbiased
one-to-one comparison of the computed mismatches is not possible.

We next investigate the implications of the mismatch on detecting a
particular model in matched-filtering analysis.  A reduction in the
match $\mathcal{M}$ is equivalent to a reduction in strain amplitude
of the exact signal $h$ (which here we assume to be given by the
waveform computed via CCE) by $\mathcal{M}h$, hence effectively
reducing the range of a GW detector by a factor of $\mathcal{M}$.  To
zeroth-order approximation, the number of detected events is proportional to
the range cubed.  A reduction in range by $\mathcal{M}$ means a
reduction of the number of detectable events by $\mathcal{M}^3$.  If
we require a loss of no more than $10\%$ of all detectable events, the
match (mismatch) between template waveform and exact signal must
therefore never go below (above) $\mathcal{M}=0.965$
($\mathcal{M}_{\rm{mis}}=3.5\times10^{-2}$)
\cite{Owen:1995tm,Damour:1997ub,Abbott:2005qm}.  This indicates that
when used as hypothetical templates in matched-filtering analysis of
the LIGO and advanced LIGO data stream, NP extraction and the
quadrupole approximation yield waveforms that are within the error
tolerance, but RWZM is generally not.

Overall, we conclude that NP extraction performs slightly worse than
the quadrupole approximation when compared to CCE.  The main reasons
for this are (i) that NP requires a higher low-frequency cut-off and
is therefore missing important low-frequency components, (ii) that NP
yields larger values for the first wave peak compared to what is
obtained with CCE or the quadrupole approximation, and (iii) that the
mismatches between NP and CCE waveforms are larger than those between
the QF and CCE waveforms.  The RWZM formalism is generally performing
the worst since it produces artificial high-frequency contributions.

\subsubsection{Variations of the Quadrupole Formula}

\begin{figure}
  \includegraphics[width=1.\linewidth]{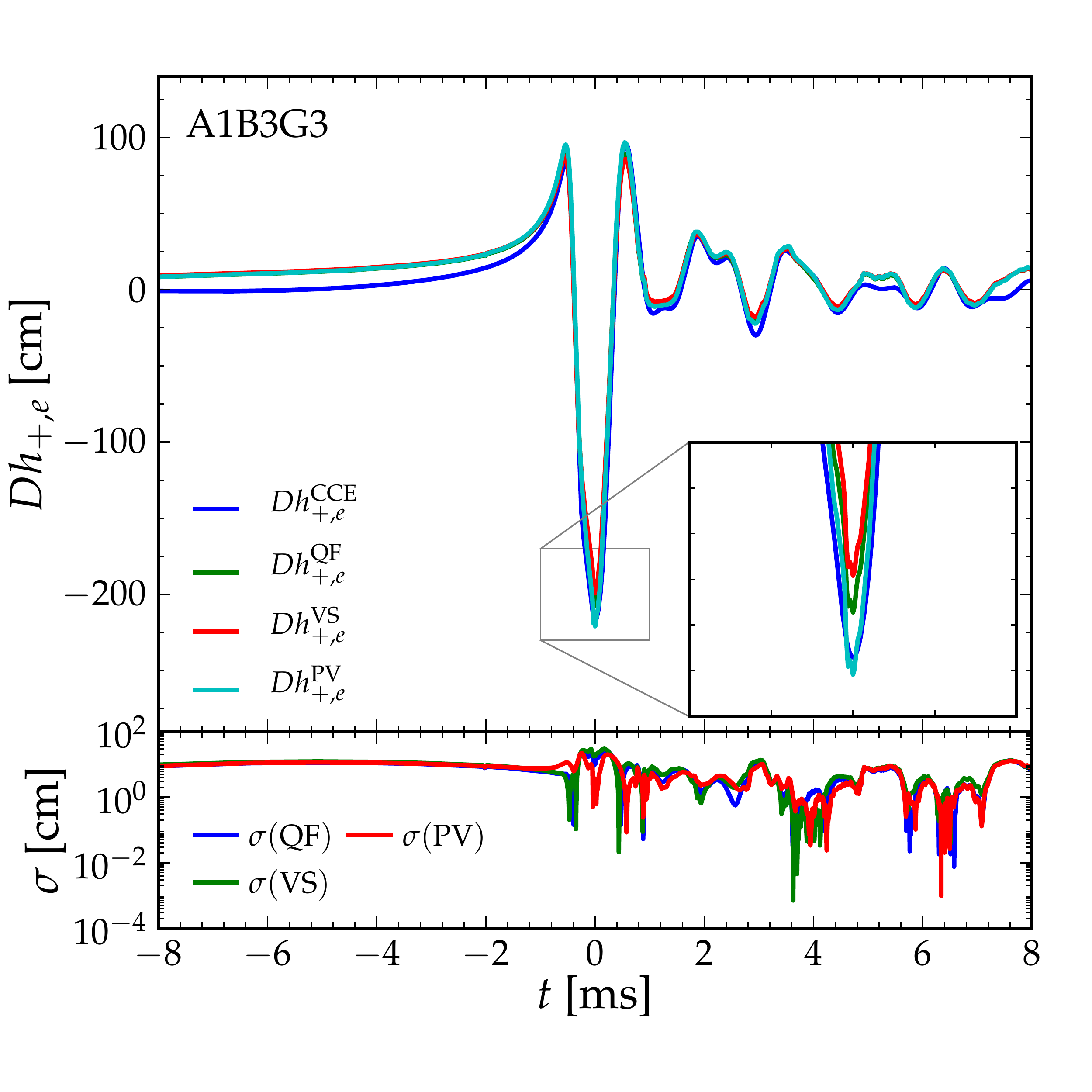}
  \vspace{-0.5cm}
  \caption{Comparison of quadrupole to CCE waveforms for model
           A1B3G3. At and immediately after core bounce, the PV
           variant leads to a marginally smaller absolute difference
           to CCE ($\sim1\%$) than any of the other QF variants.}
  \label{fig:comp_quad_A1B3G3}
\end{figure}

\begin{figure}
  \includegraphics[width=1.\linewidth]{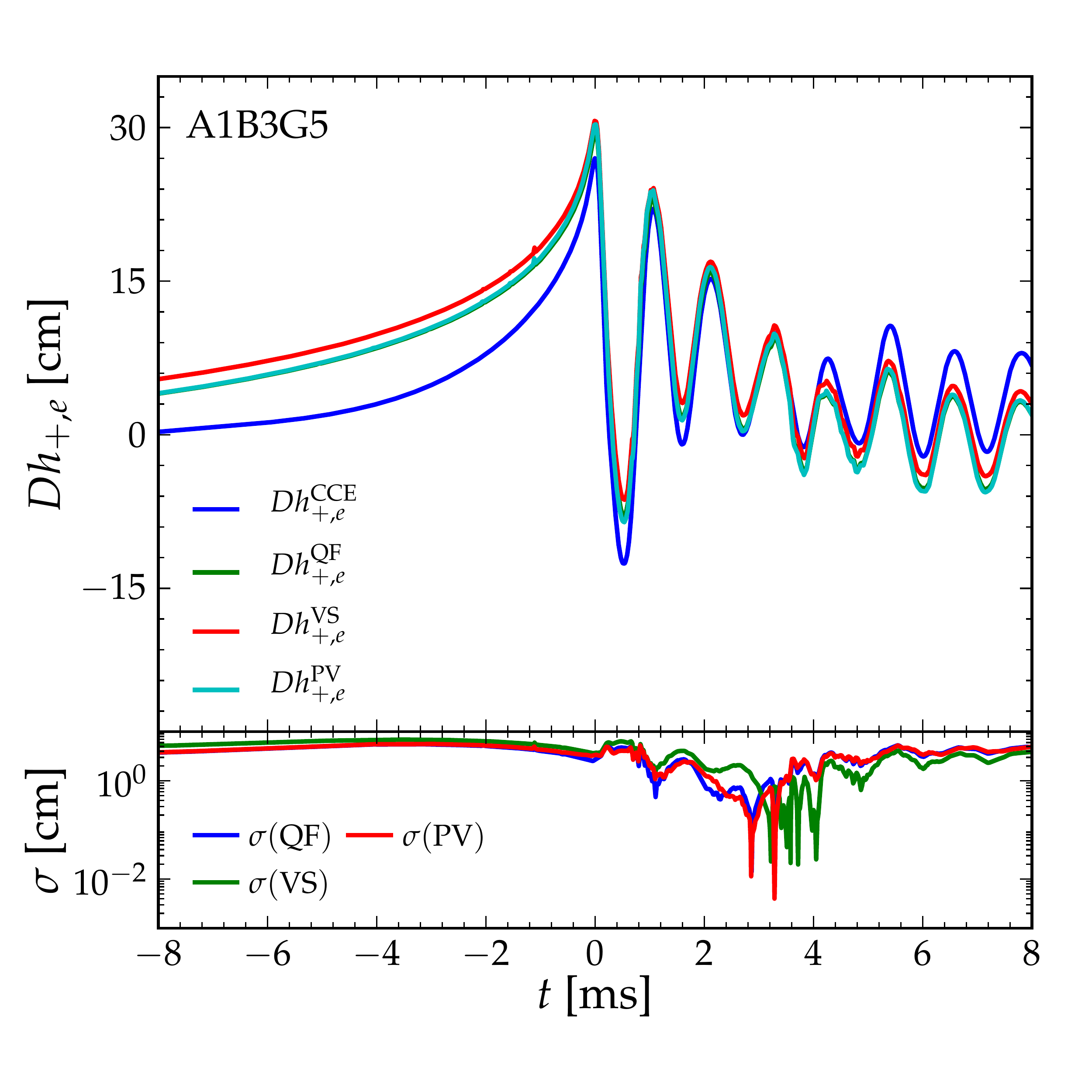}
  \vspace{-0.5cm}
  \caption{Comparison of quadrupole to CCE waveforms for model A1B3G5.
           The PV variant results in the smallest differences of the
           \textit{change} $\delta Dh_{+,e}$ between the first and second
           peaks (by $\sim2\%$) compared to CCE.  }
  \label{fig:comp_quad_A1B3G5}
\end{figure}

\begin{figure}
  \includegraphics[width=1.\linewidth]{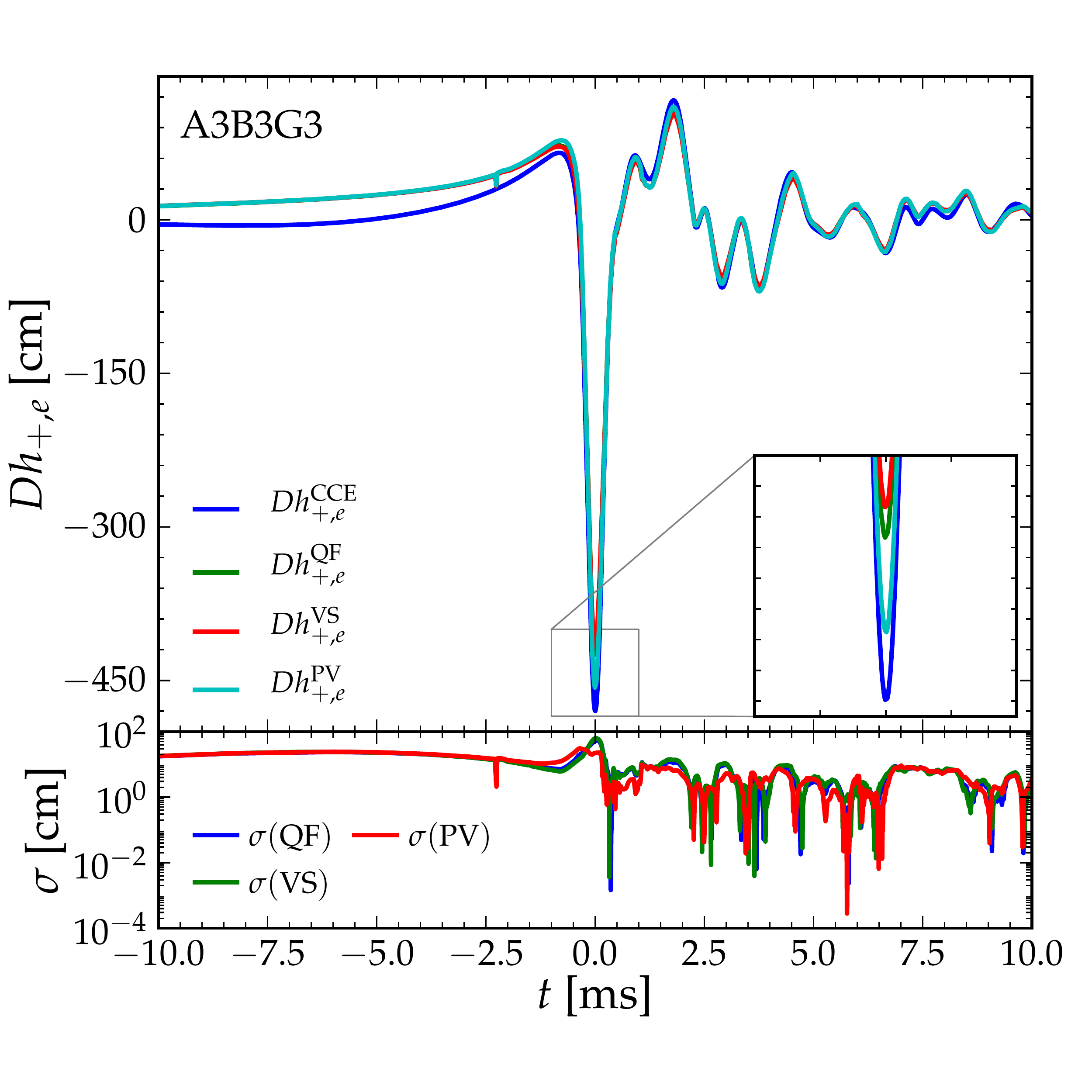}
  \vspace{-0.5cm}
  \caption{Comparison of quadrupole to CCE waveforms for model A3B3G3.
           As for model A1B3G3 (Fig.~\ref{fig:comp_quad_A1B3G3}), the PV variant
           leads to the smallest difference from CCE ($\sim5\%$).}
  \label{fig:comp_quad_A3B3G3}
\end{figure}

In Figs.~\ref{fig:comp_quad_A1B3G3}, \ref{fig:comp_quad_A1B3G5} and
\ref{fig:comp_quad_A3B3G3}, we plot waveforms for models A1B3G3,
A1B3G5 and A3B3G3, respectively, computed via the QF given in
Eq.~(\ref{eq:quad-strain}), its PV and VS variants
(Sec.~\ref{sec:quad}), and the waves as predicted by the CCE method.
In the lower panel of these figures, we plot the absolute differences
$\sigma$ from the CCE method for each variant of the QF\@.  At core
bounce, the smallest difference from CCE for model A1B3G3 is predicted
by the PV variant ($\sim1\%$ overprediction), followed by the standard
QF ($\sim5\%$ underprediction), and finally the VS variant ($\sim8\%$
underprediction).  For model A3B3G3, we measure the smallest
difference from CCE in the PV variant ($\sim5\%$ underprediction),
followed by the standard QF ($\sim11\%$ underprediction), and the VS
variant ($\sim13\%$ underprediction).  For model A1B3G5, the cut-off
of low-frequency components leads to an offset of the CCE waveform
compared to the waves obtained with the QF variants.  We therefore
compare the change in amplitude $\delta Dh_{+,e}$ between the first and
second peaks.  We find that when compared to CCE, the \emph{change}
$\delta Dh_{+,e}^{\rm PV}$ is $\sim2\%$ smaller in the PV variant,
$\sim7\%$ smaller in the standard QF, and also $\sim7\%$ smaller in
the VS variant.  Overall, the waves computed with the PV variant are
closest to the results obtained with the CCE method.  Since
the definition of the QF is ambiguous, this finding may depend on the
particular system studied and we cannot make strong general statements
in support of one or the other variant.

\subsection{Convergence}
\label{sec:convergence}

We check for convergence of our results via a resolution study of
model A3B3G3 using three different resolutions, with finest
resolutions $\Delta x=0.3\,M_\odot$ (low), $\Delta
x=0.25\,M_\odot$ (medium; our baseline resolution), and $\Delta
x=0.20\,M_\odot$ (high).

In (relativistic) hydrodynamics simulations, convergence is notoriously
difficult to analyze. The reasons are two-fold: First, the
occurrence of hydrodynamical shocks reduces the accuracy locally to
first order. In our models, a shock appears right after core bounce
and has significant impact on the order of convergence of our
scheme.  Second, our simulations are subject to some
turbulence of the fluid motion appearing soon after bounce.  
Turbulence is stochastic in nature.
Even a slight change of the resolution can result in
completely different local behavior of the fluid.  For this reason, it
is impossible to check convergence locally at each grid point.
Global quantities, however, should still be convergent. 
A sufficient global observable is the gravitational waveform and
we perform a convergence check on the waveform amplitudes.

Since we do not have an exact solution to compare with, we perform a
three-level convergence check, i.e.,~we compute the ratio of the
differences in the strain $Dh_{+,e}$ between the three resolutions
\begin{equation}
 C=\frac{|Dh^{\rm medium}_{+,e}-Dh^{\rm low}_{+,e}|}{|Dh^{\rm high}_{+,e}-Dh^{\rm medium}_{+,e}|}\,.
\end{equation}
The ratio $C$ defines the convergence \textit{factor} of the solution, and can be
translated into the order of convergence of the numerics, i.e., the convergence \textit{rate}.  Since our
lowest order of accuracy is given by first order near shocks, we expect at least first-order convergence.

\begin{figure}
  \includegraphics[width=1.\linewidth]{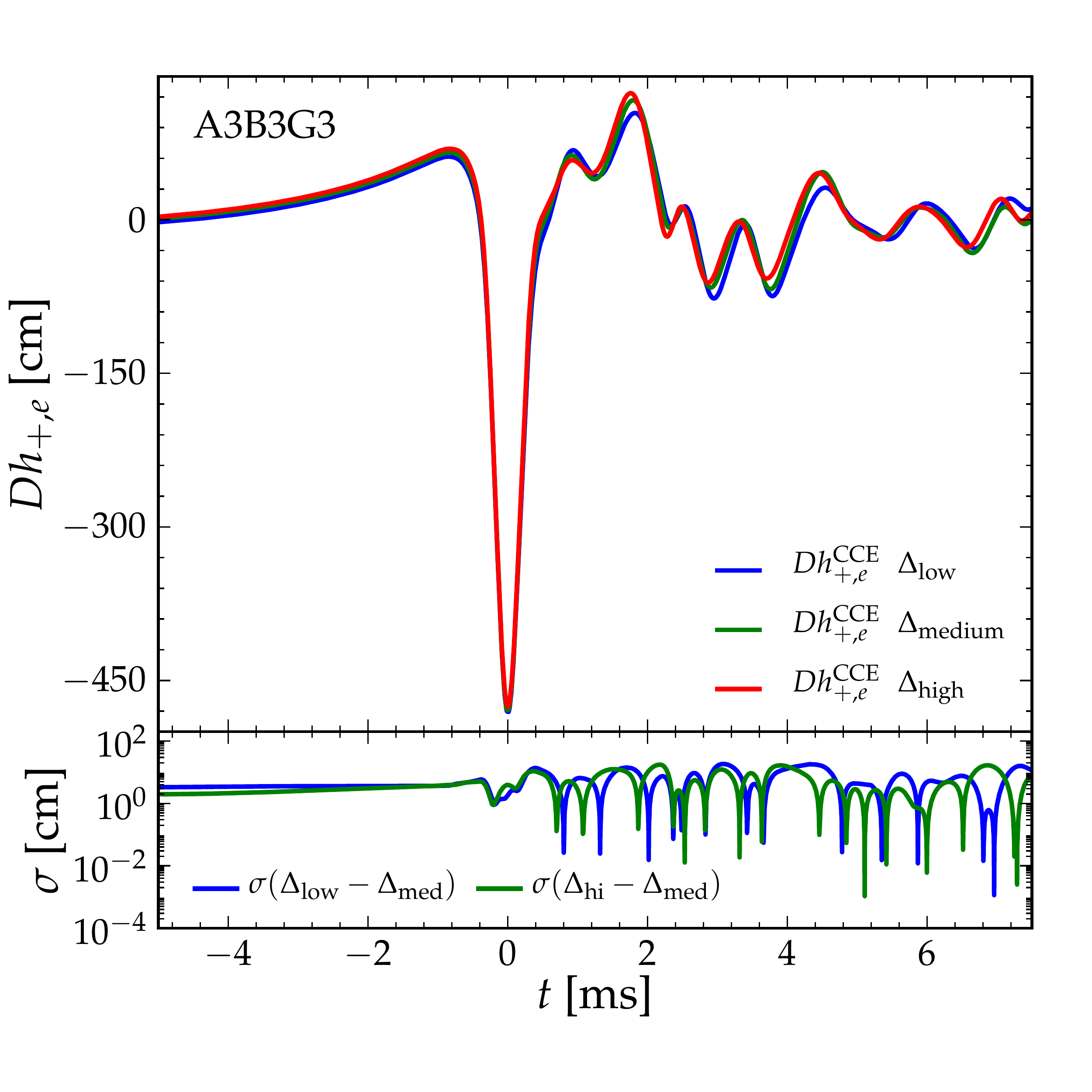}
  \vspace{-0.5cm}
  \caption{Convergence of the waveform of model A3B3G3 as computed at
    $\scri^+$ using CCE\@.  In the bottom panel, we show the absolute
    differences $\sigma$ between low and medium resolution, and high
    and medium resolutions, respectively.  Without rescaling any of
    the difference curves, we observe that they approximately line up
    at and after core bounce so that the convergence factor is simply
    given by $C\sim1$. This indicates a convergence order at and after
    core bounce of $1$.  In the prebounce phase, the difference
    between medium and high resolution is slightly smaller than the
    difference between medium and low resolutions, resulting in a
    slightly larger convergence factor.  Given our resolutions, this
    factor corresponds roughly to second-order convergence.}
  \label{fig:conv_h}
\end{figure}

Checking for convergence in the strain $D h_{+,e}$ as
computed from all extraction methods, we find a convergence factor of
$C\gtrsim 1$ prior to core-bounce and $C\sim 1$ after core bounce.
For instance, in Fig.~\ref{fig:conv_h}, we 
show a convergence plot of the waveform $D h_{+,e}^{\rm CCE}$ obtained from the CCE method\footnote{
The characteristic computational grid resolutions are scaled by the same factors as the corresponding resolutions of the Cauchy evolution.}. 
In the upper panel, we show the waveforms obtained from three different resolutions, while in the
lower panel, we show the differences between medium and low (blue curve) and high and medium resolutions (red curve).
Given our resolutions, this convergence factor corresponds to
a convergence rate between first and second order prior to core bounce which reduces to first order
after core bounce. 

We can estimate a numerical error in the medium resolution simulation by
performing a Richardson extrapolation using the measured convergence rate on the computed waveform (see e.g.~\cite{Alcubierre:2008}).
This error estimate, however, can only be applied if the convergence rate is unambiguous.
Since we measure a rate between first and second order, this is not exactly the case here.
Another measure of the numerical error in the medium resolution simulation 
is therefore directly given by
the difference between medium and high resolution simulations. This is displayed by the green curve
in the bottom panel of Fig.~\ref{fig:conv_h}.
At core bounce, we measure in the medium resolution simulation a relative numerical error of $\sim 4\%$.

\section{Summary and Conclusions}
\label{sec:summary}

We have performed a comparison study of four currently available GW
extraction techniques in the context of axisymmetric rotating stellar
core collapse.  This study is the first to succeed in extracting
GWs directly from axisymmetric core collapse spacetimes and the first
to employ the fully coordinate independent CCE extraction method for
non-vacuum spacetimes.

We have performed core collapse simulations with simplified
microphysics using a set of three representative initial
configurations leading to GW signals of varying strength and signal
morphology in quantitative agreement with what is expected from
microphysically more complete models.  In addition to having extracted
waves with variants of the standard coordinate-dependent slow-motion,
weak-field quadrupole formula, we have employed (i) the
Regge-Wheeler-Zerilli-Moncrief (RWZM) formalism, (ii) extraction based
on the Newman-Penrose (NP) scalar $\Psi_4$, and (iii)
Cauchy-characteristic extraction (CCE\@).  Of these three latter
curvature-based methods, RWZM and NP extract GWs at a finite radius
from the source, and hence, are generally prone to systematic errors
arising from (i) near-zone effects, (ii) gauge ambiguities, and (iii)
non-vanishing matter contributions.  The CCE method, on the other
hand, extracts waves gauge invariantly at future null infinity
$\mathcal{J}^+$, that is, at an infinite distance from the source
where gravitational radiation is unambiguously defined.  Hence, it is
subject only to small systematic errors due to the presence of matter
fields at the CCE world-tube locations.

An integral ingredient contributing to our success in extracting GWs
from core collapse using curvature-based methods has been the removal
of unphysical non-linear low-frequency drifts from the waveforms that
otherwise would make a proper analysis largely impossible.  This
has been achieved by the application of fixed-frequency integration
(FFI, \cite{Reisswig:2010di}) for time integration and filtering to
yield the strain $h$.

Comparing the waveforms obtained with the various extraction methods,
we make a number of observations: (i) NP- and CCE-extracted waveforms
converge with extraction and world-tube radius, respectively.  The
waveforms obtained with the RWZM formalism show spurious
high-frequency components that no other method reproduces. A number of
tests imply that the RWZM method may be less applicable to weak GW
signals, at least at the currently accessible numerical resolutions
and grid sizes.  (ii) NP extraction, CCE, and even the quadrupole
approximation, yield waveforms which agree well in phase, with
differences in the time lags between successive peaks of
$\lesssim0.05\,\rm{ms}$.  Since the RWZM formalism is contaminated by
unphysical high-frequency components, an accurate determination of the
phasing compared to the other methods is largely impossible.  (iii)
The maximum amplitudes at core bounce are different by $\sim1-7\%$ in
waveforms obtained with NP extraction and are systematically smaller
by $\sim5-11\%$ in waveforms obtained with the QF compared to the
waves obtained via CCE\@.  Accordingly, CCE yields waveforms that
result in slightly higher signal-to-noise-ratios (SNRs) ($\sim6-9\%$).
(iv) Overall, the error of the waveforms computed with the quadrupole
approximation are well within numerical errors and physical
uncertainties.  Unlike the waveforms obtained with the curvature-based
methods, the quadrupole waveforms do not suffer from low-frequency
drifts. In that respect, the quadrupole approximation is advantageous.
We also observe that the quadrupole variant using ``physical``
velocity components \cite{dimmelmeier:02a} yields waves that are
closer to those obtained via CCE\@.  However, this finding may be true
only for the core collapse case studied here and may not hold in
general.  (v) While it is unlikely that matched filtering approaches
will be used in searches for GWs from core collapse in the near
future, we have nevertheless computed GW template mismatches, a
measure for the detectability of differences between waveforms.  We
find that when used in hypothetical matched-filtering GW searches,
waveforms from NP extraction, CCE, and the QF would lead to the
detection of the same model, while the waveforms computed with the
RWZM formalism would generally not.

There are two major drawbacks of our current work: (i) The
curvature-based methods assume vacuum at the extraction spheres and
world-tube locations. Hence, we must, in principle, extract at 
very large radii where the stress-energy tensor is zero.
This, however, is currently not possible, since the collapsing star
extends over the entire computational grid and larger grids are
computationally prohibitive.  
(ii) All curvature-based methods yield waveforms with unphysical
low-frequency drifts, requiring removal by spectral cut-off via
FFI. This is particularly problematic in models with physical content
below $\sim100\,\rm{Hz}$.
A possible improvement of the low-frequency behavior could be achieved
by the inclusion of matter terms in the CCE method, or alternatively,
by enlarging the simulation domain such that the extraction takes
place outside of the star and in pure vacuum.  The latter could be
efficiently achieved by employing multiblock techniques that cover the
wavezone by a set of spherical grids \cite{zink:08b}.

Finally, we point out that we have considered only the GW signal from
rotating core collapse and bounce in this first study using
curvature-based GW extraction from core collapse spacetimes.  While
our results may transfer to other GW emission processes in core
collapse, this is by no means guaranteed. Further work will be needed
to adress curvature-based GW extraction also from postbounce
convection and the standing accretion shock instability, protoneutron
star pulsations, rotational instabilities, and black hole formation.


\acknowledgments

We are happy to acknowledge helpful exchanges with E.~Abdikamalov,
P.~Ajith, A.~Burrows, N.~T.~Bishop, P.~Cerd\'a-Dur\'an, P.~Diener,
H.~Dimmelmeier, J.~Kaplan, E.~O'Connor, E.~Pazos, D.~Pollney,
E.~Seidel, R.~O'Shaughnessy, K.~Thorne, and S.~Teukolsky.  This work is supported by the
Sherman Fairchild Foundation and by the National Science Foundation
under grant numbers AST-0855535, OCI-0721915, PHY-0904015,
OCI-0905046, and OCI-0941653.  CDO and CR wish to thank Chris Mach for
support of the group servers at TAPIR on which much of the code
development and testing was carried out.  Results presented in this
article were obtained through computations on the Caltech compute
cluster ``Zwicky'' (NSF MRI award No.\ PHY-0960291), on the NSF
Teragrid under grant TG-PHY100033, on machines of the Louisiana
Optical Network Initiative under grant loni\_numrel05, and at the
National Energy Research Scientific Computing Center (NERSC), which is
supported by the Office of Science of the US Department of Energy
under contract DE-AC03-76SF00098.  US acknowledges support from the
Ram{\'o}n y Cajal Programme of the Spanish Ministry of Education and
Science, from FCT-Portugal through PTDC/FIS/098025/2008 and
allocations through the TeraGrid Advanced Support Program under grant
PHY-090003 at NICS and the Centro de Supercomputaci\'on de Galicia
(CESGA, project number ICTS-2009-40).


\bibliography{bibliography/gw_references,bibliography/sn_theory_references,bibliography/grb_references,bibliography/stellarevolution_references,bibliography/methods_references,bibliography/uli,bibliography/reisswig,bibliography/gw_data_analysis_references,bibliography/NSNS_NSBH_references,bibliography/misc_references,bibliography/cs_hpc_references,bibliography/numrel_references,bibliography/publications-schnetter,bibliography/bh_formation_references}

\end{document}